\documentclass[fleqn,usenatbib]{mnras}

\usepackage{newtxtext,newtxmath}

\usepackage[T1]{fontenc}

\DeclareRobustCommand{\VAN}[3]{#2}
\let\VANthebibliography\thebibliography
\def\thebibliography{\DeclareRobustCommand{\VAN}[3]{##3}\VANthebibliography}

\usepackage{graphicx}	
\usepackage{amsmath}	
\usepackage{bm}

\usepackage{bm}
\usepackage{color}
\usepackage{ulem}
\usepackage{xspace}
\usepackage{float}
\usepackage{graphicx}
\usepackage{epstopdf}
\usepackage{CJKutf8}
\usepackage{multirow}

\ttfamily
\definecolor{mygray}{gray}{0.6}
\definecolor{orange}{rgb}{1.0, 0.4, 0.0}
\definecolor{easygreen}{rgb}{0.1, 0.7, 0.1}
\definecolor{magenta}{rgb}{0.858, 0.188, 0.478}
\definecolor{revise}{RGB}{255, 0, 0 }

\newcommand{\fg}[1]{Fig.~\ref{fig:#1}}

\newcommand{\eq}[1]{Eq.~(\ref{eq:#1})\xspace}
\newcommand{\Eq}[1]{Equation~(\ref{eq:#1})\xspace}

\newcommand{\tb}[1]{Table~\ref{tab:#1}\xspace}
\newcommand{\se}[1]{Sect.~\ref{sec:#1}\xspace}

\newcommand{\Ap}[1]{Appendix~\ref{sec:#1}\xspace}

\newcommand{\stalling}{\texttt{Braking}\xspace}
\newcommand{\migrating}{\texttt{Migrating}\xspace}

\usepackage{ulem}
\makeatletter
\def\uwave{\bgroup \markoverwith{\lower3.5\p@\hbox{\sixly \textcolor{red}{\char58}}}\ULon}
\font\sixly=lasy6 
\makeatother

\title[Resonant planets and their implications]{When, where, and how many planets end up in first-order resonances?}

\author[Shuo Huang \& Chris W. Ormel]{
Shuo Huang (\begin{CJK*}{UTF8}{gbsn}黄硕\end{CJK*}),$^{1,2}$\thanks{E-mail: huangs20@mails.tsinghua.edu.cn}
Chris W. Ormel,$^{1}$\thanks{E-mail: chrisormel@tsinghua.edu.cn}
\\
$^{1}$ Department of Astronomy, Tsinghua University, 30 Shuangqing Rd, 100084 Beijing, China \\
$^{2}$ Leiden Observatory, Leiden University, P.O. Box 9513, 2300 RA Leiden,
The Netherlands}

\date{Accepted XXX. Received YYY; in original form ZZZ}

\pubyear{2022}

\begin{document}
\label{firstpage}
\pagerange{\pageref{firstpage}--\pageref{lastpage}}
\maketitle

\begin{abstract}
The theory of Type~I migration has been widely used in many studies.  
Transiting multi-planet systems offer us the opportunity to examine the consistency between observation and theory, especially for those systems harbouring planets in Mean Motion Resonance (MMR). The displacement these resonant pairs show from exact commensurability provides us with information on their migration and eccentricity-damping histories. Here, we adopt a probabilistic approach, characterized by two distributions -- appropriate for either the resonant or non-resonant planets -- to fit the observed planet period ratio distribution. With the Markov chain Monte Carlo (MCMC) method, we find that ${\approx}15\%$ of exoplanets are in first order ($j+1{:}j$) MMRs, the ratio of eccentricity-to-semi-major axis damping is too high to allow overstable librations and that the results are by-and-large consistent with Type-I migration theory. In addition, our modeling finds that a small fraction of resonant pairs is captured into resonance during migration, implying late planet formation (gas-poor). Most of the resonant pairs park themselves at the migration barrier, indicating early planet formation (gas-rich). Furthermore, after improving the criterion on two-body resonant trapping, we obtain an upper limit of the disc surface density at the time the planets are locked in resonance. 


\end{abstract}




\section{Introduction}
Since the first discovery of exoplanets around solar-type stars \citep{MayorQueloz1995}, the number of exoplanets has ballooned in the last three decades, exceeding 5\,200 as of the present day. It is therefore appropriate to conduct population-level analyses to examine planet formation theories \citep{MordasiniEtal2015, ZhuDong2021}. When independent mass and radius measurements are available, planet bulk density and their composition can be inferred \citep{FortneyEtal2007, SeagerEtal2007, PiauletEtal2022}, with which their mass accretion history and post-formation evolution can be constrained. The core accretion model successfully predicted the so-called "planet dessert" \citep{IdaLin2004}, which refers to the paucity of planets with tens of Earth-mass within 3~au. The "radius valley" \citep{FultonEtal2017} is manifested at a planet radius ${\sim}2R_\oplus$, which has been attributed to photoevaporation-driven mass loss \citep{OwenWu2013, OwenWu2017}, planet formation location with respect to snow line \citep{LuquePalle2022, IzidoroEtal2022}, or core-powered mass loss \citep{GinzburgEtal2018}.

The physical principles underlying migration of low-mass planets in gaseous discs (Type~I migration) have long been established \citep{GoldreichTremaine1979, LinPapaloizou1979}. The total torque exerted on planets by the surrounding disc is typically negative, resulting in inward planet migration on time-scales shorter than the disc lifetimes \citep[e.g.,][]{Ward1997, TanakaEtal2002, RibasEtal2014, WinterEtal2019}. However, the direction of migration can be reversed at special locations where conditions materialize that render a net positive torque, resulting in migration traps. These locations include the region where the horseshoe saturates, i.e when the (positive) co-rotation torque compensates the (negative) Lindblad torque \citep{GoldreichTremaine1980, Ward1991, PaardekooperEtal2010, PaardekooperEtal2011}, the disc inner edge \citep{LiuEtal2017, RomanovaEtal2019, AtaieeKley2021} where the torque becomes one-sided, and the regions where the disc switches from optically thin to optically thick ($0.1-1$ au) \citep{MassetEtal2006}. In addition, in the pebble accretion paradigm, the infalling dust can efficiently induce (positive) thermal torque onto the planets \citep{Benitez-LlambayEtal2015, Masset2017, GuileraEtal2019, GuileraEtal2021}. 

Although the present close-in positions of exoplanets indirectly hint at planet migration, it is hard to quantitatively test the theory based only on {single-planet systems}. Instead, {multi-planet systems, especially those with} planet pairs in Mean Motion Resonance (MMR) leave richer dynamical imprints against which the theory can be tested \citep{SnellgroveEtal2001, PapaloizouSzuszkiewicz2005}. It is likely that such resonant architecture results from migration in a gas-rich environment (i.e., the disc) as energy dissipation is needed to trap planets in resonance \citep{TerquemPapaloizou2007, RaymondEtal2008, Rein2012, Batygin2015i}. {One famous example is PDS 70, which harbours two directly imaged near-resonance planets in its protoplanetary disc \citep{BaeEtal2019, BenistyEtal2021}. Besides, a chain of planets in resonance might sculpt the asymmetry disc structure in HD 163296 \citep{IsellaEtal2018, Garrido-DeutelmoserEtal2023}. } In particular, there tends to be an excess of systems with planets' period ratio just wide of commensurability \citep{FabryckyEtal2014, SteffenHwang2015}, indicating that a certain fraction of planet pairs are truly in resonance. 

{Formally, two planets are said to be in $(j+1):j$ resonance if at least one of their resonance angles ($\phi_\mathrm{1,2}=(j+1)\lambda_2-j\lambda_1-\varpi_\mathrm{1,2}$, with $\lambda_i$ the mean longitudes and $\varpi_i$ the longitude of pericentres) librate around a fixed value. However, the values of the resonance angles are poorly constrained because it is hard to constrain $\varpi_i$ for near-circular orbits. We therefore turn our attention to their period ratios}
and define the dimensionless parameter
\begin{equation}
    \Delta = \frac{P_2}{P_1}-\frac{j+1}{j}
\end{equation}
to measure the offset of the period ratio away from a first-order $(j+1):j$ commensurability. Here, $P_1$ is the period of the inner planet and $P_2$ that of the outer. If two planets are in a $(j+1):j$ resonance, the offset $\Delta$ must be close to zero. 
\cite{Xie2014} and \cite{RamosEtal2017} emphasize that the exact value of $\Delta$ is determined by migration and eccentricity damping, linking the observed quantity $\Delta$ to planet migration \citep{CharalambousEtal2022}. 
It offers us an opportunity to examine planet-disc interaction histories through planets in MMR in multi-planet systems. 
The migration history of such specific multi-planet systems like TRAPPIST-1 \citep{GillonEtal2017, LugerEtal2017, HuangOrmel2022i}, K2-24 \citep{PetiguraEtal2018, TeyssandierLibert2020} and TOI-1136 \citep{DaiEtal2022}, can be reconstructed. 

Yet, most exoplanets are obviously not in resonance because of their large offsets $\Delta$. Various scenarios have been proposed to explain the overall observed non-resonant planetary architecture statistically. These include dynamical instability \citep{IzidoroEtal2017, IzidoroEtal2021}, disc winds \citep{OgiharaEtal2018}, \textit{in situ} formation of sub-Netunes \citep{DawsonEtal2015, ChoksiChiang2020}, planetesimal scattering \citep{ChatterjeeFord2015, GhoshChatterjee2022}, stellar tides \citep{LithwickWu2012, DelisleLaskar2014, Xie2014, SanchezEtal2020}, and stochastic forces \citep{ReinPapaloizou2009, GoldbergBatygin2022}. 
However, many of these works introduce additional free parameters in order to match the only observed quantity -- the period ratio distribution. Overfitting may occur. One way to improve on this is to include more observational quantities. For example, \cite{GoldbergBatygin2022} and \cite{ChoksiChiang2022} find that by additionally accounting for the TTV signatures of hundreds of Kepler planets, a laminar disc alone cannot reproduce the observed TTV features. Either additional planets (perturbers) are required \citep{ChoksiChiang2022} or their birth proto-discs are turbulent \citep{GoldbergBatygin2022}.


In this work, we introduce a statistical approach to study all transiting exoplanets and try to answer when, where, and how many planets are captured in resonance. The observed planetary radii, orbital periods, and host stellar masses are taken into account. By assuming that the period ratio distribution is characterized by two distributions, representing both resonant and non-resonant planets, the probability that a planets pair is in resonance is evaluated. Whether the migration and eccentricity damping is consistent with migration theory is determined. In addition, our approach allows us to constrain the timing and the location, of the resonance trapping, which implies the pathway of planet formation.



The paper is structured as follows: We first introduce our statistical model in \se{methodology}. Using the MCMC method, we constrain the relation between eccentricity damping time-scale and migration time-scale in \se{mcmc_analysis}. Along with the resonance trapping criterion that we improve on (\se{new_criterion}), we address when, where, and how many planets are in resonance \se{implication}. The discussion and conclusion of this study are presented in \se{discussion} and \se{conclusion}, respectively.


\section{Methodology}
\label{sec:methodology}
In this section, we {describe our disc model and migration (Type I)}, and construct the likelihood function needed for the Monte Carlo Markov Chain (MCMC) simulations of \se{mcmc_analysis}.
Since our migration model is linear {with planet mass} (Type I), we focus on the planets with relatively low planet-to-star mass ratios {which are not likely to open a gap when they are in the protoplanet disc}. We assume that the relevant disc quantities follow power-law distribution (\se{disc_model}). We describe the migration model in \se{migration}. The equilibrium dynamics of planets trapping in resonance are described in \se{resonance}. The masses of observed planets are calculated from a mass-radius (M-R) relationship. Its prescription is given in \se{MR-relation}. The total log-likelihood function we construct for the MCMC is detailed in \se{MCMC_model}. 

\subsection{Disc model}
\label{sec:disc_model}
{As most observed transiting planets are located within $\sim$1 au of their host star, we will describe the possible structures of the inner disc.} We assume that the gas surface density always follows a power-law distribution:
\begin{equation}
\label{eq:power_law_sigma}
    \Sigma(r) = \Sigma_\mathrm{1au} \left(\frac{r}{1\,\mathrm{au}}\right)^{s},
\end{equation}
where $\Sigma_\mathrm{1au}$ is the gas surface density at $1\,\mathrm{au}$ and $s$ is its slope. The gas aspect ratio also follows a power-law distribution:
\begin{equation}
\label{eq:power_law_h}
    h(r) = h_\mathrm{1au} \left(\frac{r}{1\,\mathrm{au}}\right)^{q}
\end{equation}
where $h_\mathrm{1au}$ is the gas aspect ratio at $1\,\mathrm{au}$ and $q$ is its slope. Different assumptions about the disc structure, e.g., heating mechanisms, result in distinct values of $s$, $q$, $\Sigma_\mathrm{1au}$, and $h_\mathrm{1au}$. Typically, the inner disc is optically thick and the main heating energy comes from viscous dissipation \citep{RudenLin1986}, while the outer disc is optically thin and stellar irradiation mainly heat the disc onto its surface layer \citep{ChiangGoldreich1997}. 

{For discs} dominated by stellar radiation, we make use of the disc structure from \citet{LiuEtal2019}. The gas surface density is: 
\begin{equation}
        \Sigma_{\mathrm{g,irr}}= 250\left(\frac{\dot{M}_g}{10^{-8}M_\odot /\mathrm{yr}}\right)\left(\frac{M_\star}{M_\odot}\right)^\frac{9}{14}\left(\frac{L_\star}{L_\odot}\right)^{-\frac{2}{7}}
        \left(\frac{r}{\mathrm{au}}\right)^{-\frac{15}{14}} \mathrm{g\,cm}^{-2}
\end{equation}
where $\dot{M}_g$ is stellar accretion rate and $L_\star$ is the star luminosity. The aspect ratio is:
\begin{equation}
\label{eq:h_g_irr}
    \begin{aligned}
        h_{\mathrm{g,irr}}=& 0.0245\left(\frac{M_\star}{1M_\odot}\right)^{-4/7}\left(\frac{L_\star}{1L_\odot}\right)^{1/7}\left(\frac{r}{1\,\mathrm{au}}\right)^{2/7}.
    \end{aligned}
\end{equation}
For stars of mass between $0.43M_\odot$ and $2M_\odot$, which covers most of our star sample, the mass-luminosity relation is well represented by ${L_\star}/{L_\odot}=\left({M_\star}/{M_\odot}\right)^4$ \citep{Duric2004}.
Therefore, the disc aspect ratio simplifies to:
\begin{equation}
    h_{\mathrm{g,irr}}=0.0245\left(\frac{r}{1\,\mathrm{au}}\right)^{2/7}
\end{equation}
independent of stellar mass. 

If the {inner} disc is dominated by viscous heating, its temperature structure is highly related to the viscous accretion rate and opacity. Following \cite{LiuEtal2019}, $s$ and $q$ are taken to be $-3/8$ and $-1/16$, while $\Sigma_\mathrm{1au}$ and $h_\mathrm{1au}$ are not specified. 

\subsection{Type I migration}
\label{sec:migration}
In the Type I migration regime, planet migration is the result of a net torque {$\Gamma_\mathrm{net}$ consisting of the} Lindblad \citep{Ward1986,Ward1997}, corotation \citep{GoldreichTremaine1979,Ward1992} and thermal torques \citep{Benitez-LlambayEtal2015, Masset2017, GuileraEtal2019, GuileraEtal2021}, etc. Usually, the net torque is negative and the planet migrates inward. In the Type~I limit the migration speed is proportional to disc mass and planet mass. In the limit of a locally isothermal disc, which implies that temperature is a function of radius only, $T(r)$, the type I migration time-scale for the $i$-th planet at distance $r_i$ is:
\begin{equation}
    \label{eq:migration}
    \tau_{a_i}
    = \frac{L_i}{\Gamma_\mathrm{net}} 
    =\frac{\gamma_I \tau_{w_i}}{h(r_i)^2}, 
\end{equation}
where $L_i$ is the angular momentum of the planet, $\gamma_I=2.7-1.1s$ \citep{TanakaEtal2002} is the Type I migration prefactor \citep{D'AngeloLubow2010}, and $h(r_i)$ is the disc gas aspect ratio at $r_i$. The characteristic time of the orbital evolution \citep{TanakaWard2004} is:
\begin{equation}
    \tau_{w_i}=\frac{1}{\mu_i}\frac{M_\star}{\Sigma(r_i)r_i^2}\frac{h(r_i)^4}{\Omega_K(r_i)},
\end{equation}
where $\mu_i$ is the mass ratio of the $i$-th planet over its host star and $\Omega_K(r_i)=\sqrt{GM_\star/r_i^3}$ is the Keplerian angular velocity at distance $r_i$. 
The eccentricity damping rate is proportional to the local surface density and planet mass. It is given by:
\begin{equation}
    \label{eq:te}
    \tau_{e_i}=\frac{C_e \tau_{w_i}}{0.78}
    = \frac{C_e}{0.78\gamma_I} h^2 \tau_a
    ,
\end{equation}
where $C_e$ stands for eccentricity damping efficiency. Although \cite{CresswellNelson2008} gives $C_e\approx1$, lower values are needed in other studies to reproduce specific systems. TRAPPIST-1 planets demand $C_e\approx0.1$ \citep{HuangOrmel2022i} {and K2-24 requires $C_e\approx0.28$} \citep{TeyssandierLibert2020}. A recently discovered ${\sim}100\,\mathrm{Myr}$ old exoplanet system TOI-1136, on the other hand, suggest $C_e\sim10$ \citep{DaiEtal2022}. We are therefore agnostic about the value of $C_e$, which value we intend to constrain through our MCMC fitting.

\subsection{Dynamics of resonance trapping}
\label{sec:resonance}
Resonance trapping is a natural outcome of convergent disc migration and eccentricity damping, especially for first-order resonances $(j+1){:}j$. When two planets are in first-order resonance, their eccentricities and period ratio librate around their equilibrium values. Such equilibrium has been studied by \cite{GoldreichSchlichting2014} and \cite{TerquemPapaloizou2019}. They both give the equilibrium eccentricity (the eccentricity where tidal damping equals resonant excitation) of the inner planet: 
\begin{equation}
    \label{eq:e_{1,eq}}
    e_\mathrm{1,eq}^2=\frac{\tau_{e_1}/\tau_{a_2}-\tau_{e_1}/\tau_{a_1}}{2(j+1)\left(1+\frac{j}{j+1}\frac{\mu_1}{\alpha \mu_2}\right)\left[1+\frac{\mu_1}{\alpha \mu_2}\left(\frac{j}{j+1}\right)^2\left(\frac{f_2'}{f1}\right)^2\frac{\tau_{e_1}}{\tau_{e_2}}\right]}
\end{equation}
where $\alpha$ is ratio of the inner-to-outer semi-major axis . Here, and in the following, the subscription '1' stands for the inner planet and '2' for the outer planet. The relationship between the eccentricities of the inner and outer planets is:
\begin{equation}
    \label{eq:e1e2}
    \frac{e_2^2}{e_1^2}=\left(\frac{\mu_1}{\alpha \mu_2}\frac{j}{j+1}\frac{f_2'}{f_1}\right)^2,
\end{equation}
where $f_1$ and $f_2'$ are coefficients tabulated in \citet[][their Table A1]{TerquemPapaloizou2019}. The equilibrium value for the offset from exact resonance is:
\begin{equation}
    \label{eq:commensurability}
    \Delta_\mathrm{eq} = -f_2'\mu_1 \frac{1}{j} \frac{1}{e_\mathrm{2,eq}}.
\end{equation}
At this distance, the resonance repulsion equals the Type-I inward migration.

In order to calculate the period ratios for planets in resonance, we need to know what their equilibrium eccentricities are. This calculation can be done only after the values of ${\tau_{e_1}}/{\tau_{e_2}}$, ${\tau_{e_1}}/{\tau_{a_2}}$ and ${\tau_{e_1}}/{\tau_{a_1}}$ are known. Combining the disc model and migration model, we have
\begin{equation}
\label{eq:te1_te2}
    \frac{\tau_{e_1}}{\tau_{e_2}}=\frac{\mu_2}{\mu_1} \left(\frac{r_1}{r_2}\right)^{4q-s-0.5}
\end{equation}
and
\begin{equation}
\label{eq:te1_ta2}
    \frac{\tau_{e_1}}{\tau_{a_2}} = \frac{\tau_{e_2}}{\tau_{a_2}} \frac{\tau_{e_1}}{\tau_{e_2}}=\frac{C_e h(r_2)^2}{0.78 \gamma_I}\frac{\mu_2}{\mu_1} \left(\frac{r_1}{r_2}\right)^{4q-s-0.5}
\end{equation}
where $s$ and $q$ are gas surface density and aspect ratio gradient.

For the eccentricity-to-semi-major axis damping of the inner planet, we distinguish it between two cases:
\begin{enumerate}
    \item \underline{\migrating pair}. If resonances are formed during migration and the ambient disc disperses before the planet pairs reach a migration barrier, planet migration and eccentricity damping follow \eq{migration} and \eq{te}. Therefore, 
\begin{equation}
\label{eq:te1_ta1_m}
    \frac{\tau_{e_1}}{\tau_{a_1}}=\frac{C_e h(r_1)^2}{0.78 \gamma_I}
\end{equation}
    It requires that the outer planet migrates faster than the inner planet to guarantee convergent migration. 
    
    \item \underline{\stalling pair}. On the other hand, if resonances are formed before/after the inner planet's migration is halted by a barrier (could be disc inner edge or the radius where reverse migration occurs), there is no net torque on the two planets. Angular momentum conservation gives $\tau_{a_1}/\tau_{a_2}=-\sqrt{\mu_1 r_1/\mu_2 r_2}$. Hence,
    \begin{equation}
    \begin{aligned} 
        \label{eq:te1_ta1_s}
        \frac{\tau_{e_1}}{\tau_{a_1}}
        & =-\frac{C_e h(r_2)^2}{0.78 \gamma_I} \left(\frac{\mu_2}{\mu_1}\right)^{\frac{3}{2}} \left(\frac{r_1}{r_2}\right)^{4q-s-1} \\
        & = -\frac{C_e h(r_1)^2}{0.78 \gamma_I} \left(\frac{\mu_2}{\mu_1}\right)^{\frac{3}{2}} \left(\frac{r_1}{r_2}\right)^{2q-s-1}
    \end{aligned}
    \end{equation}
\end{enumerate}
For the outer planet ${\tau_{e_2}}/{\tau_{a_2}}={C_e h(r_2)^2}/{0.78 \gamma_I}$ is always true. It is also apparent that the resonant equilibrium does not depend on disc mass ($\Sigma_\mathrm{1au}$) but on power law indices and aspect ratio ($s$, $q$, and $h_\mathrm{1au}$). 

\begin{figure}
    \centering
    \includegraphics[width=\columnwidth]{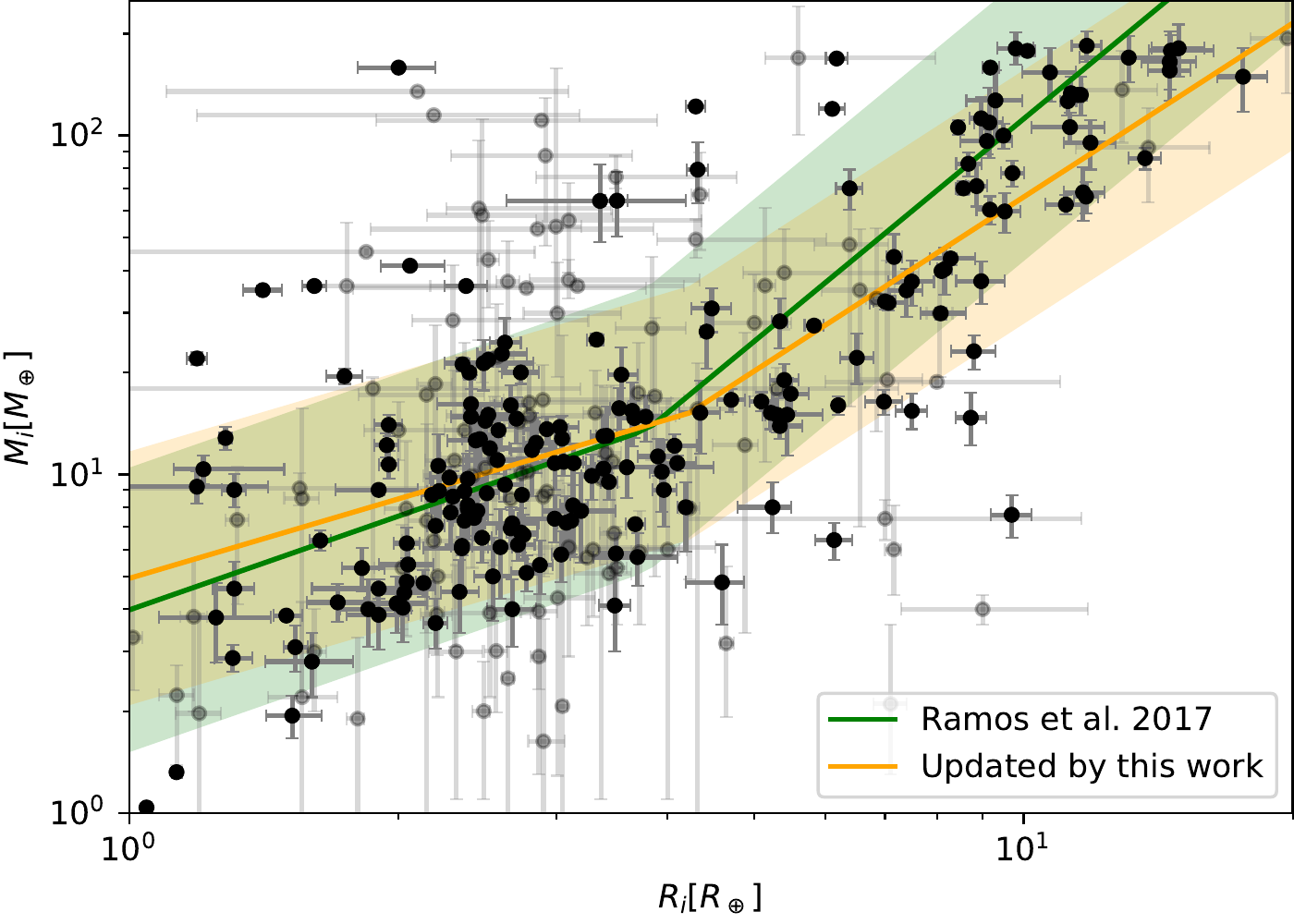}
    \caption{
    Mass-radius (M-R) relationship for exoplanets. 
    Planets are selected based on the data of transiting planets in the NASA Exoplanet Archive. Only planets with mass lower than $200M_{\earth}$, radii smaller than $20R_{\earth}$ and periods longer than $5$ days are included in the sample.
    Planets' masses and radii are indicated by black dots, with 1$\sigma$ error bars.
    The green colour indicates the fit result from \citet{RamosEtal2017}, orange colour is the updated fit result by this work (\eq{MRfitting}). The values within their 1$\sigma$ dispersion ($\sigma_\mathrm{m}=0.374$ for our fit) are contained within the light green and orange region respectively.
    }
    \label{fig:fitting}
\end{figure}

\subsection{Mass-radius relations}
\label{sec:MR-relation}
Planets' masses are also needed to calculate the equilibrium period ratios in resonance. However, most transiting planets have poorly constrained masses compared to their radii. 

If the planet mass is not yet constrained from e.g., Transit Timing Variation \citep[TTV][]{AgolEtal2005} or Radial Velocity (RV), we then obtain the planet mass using a mass-radius relation. The planet sample for fitting the mass-radius relation is based on the data of transiting planets from the NASA Exoplanet Archive\footnote{\url{https://exoplanetarchive.ipac.caltech.edu}}. Planets with masses lower than $200M_{\earth}$, radii smaller than $20R_{\earth}$, or periods longer than $5$ days {(see \se{sample})} are selected while planets with  periods shorter than $5$ days are excluded.

Our fitting approach is identical to \cite{RamosEtal2017}. They use a broken power law expression, which fits two different power law relations for larger bodies ($R>R_\mathrm{crit}$) and smaller bodies ($R<R_\mathrm{crit}$). 
\begin{equation}
    \log_{10}\left(\frac{\overline{M}_i}{M_{\earth}}\right)=
    \left\{\begin{array}{ll}
        a+b\log_{10}\left(\frac{R_i}{R_{\earth}}\right) & \text{if } R_i\le R_\mathrm{crit}\\
        c+d\log_{10}\left(\frac{R_i}{R_{\earth}}\right) & \text{if } R_i>R_\mathrm{crit}
    \end{array}
    \right.
    \label{eq:MRfitting}
\end{equation}
where $M_i$ and $R_i$ is the mass and radius of the $i$-th planet. We make use of the Maximum Likelihood Estimator (MLE) to maximize a Gaussian likelihood centered at \eq{MRfitting}. The critical radius $R_\mathrm{crit}$ is also fitted. It estimates that: $a=0.69$, $b=0.78$, $c=0.11$, $d=1.7$, $R_\mathrm{crit}=4.23R_{\earth}$ and the corresponding dispersion $\sigma_m=0.374$. The best-fitting relation is shown in \fg{fitting} in comparison with the relation fitted by \cite{RamosEtal2017}. They do not differ significantly. The estimated value for $R_\mathrm{crit}$ is consistent with \citet{TeskeEtal2021} who suggest a single power law relation for planets with $R<3.25R_\oplus$. 

\Eq{MRfitting} allows us to calculate the planet mass and therefore the resonance offset (\eq{commensurability}). Moreover, the log-normal dispersion in planet mass enables us to check our model consistency. 
The reason is that any uncertainty in the mass, will propagate, through \eq{e_{1,eq}} and \eq{commensurability}. Therefore, we expect that the obtained value for $\sigma_\Delta$ from MCMC fitting is similar to, or exceeds, $\sigma_m$.


The log-normal dispersion in planet mass significantly simplifies our analysis. There are arguably more sophisticated forms of M-R relationships, e.g., the one given by \citet{WolfgangEtal2016} and improved by \citet{TeskeEtal2021}. However, they assume planet mass follows a normal dispersion instead of a log-normal. In that case, the resulting $\Delta$ would follow a complicated form of the ratio distribution\footnote{If variables $X$ and $Y$ follow a dependent (independent) normal distribution with nonzero mean values, the new variable $Z=X/Y$ follows correlated (uncorrelated) non-central normal ratio distribution \citep{Hinkley1969, HayyaEtal1975}.}.

\subsection{A statistical model of resonant and non-resonant planets}
\label{sec:MCMC_model}
We define the posterior distribution as:
    $
    p(\bm{\theta}|\Delta_{\mathrm{obs},k},\bm{X}_k)= p(\Delta_{\mathrm{obs},k}|\bm{\theta},\bm{X}_k)p(\bm{\theta}|\bm{X}_{\mathrm{obs},k}).
    $
If the disc structure is not specified {(without knowing the specific values of $s$, $q$ in \eq{power_law_sigma} and \eq{power_law_h})}, the unknown model parameters are $\bm{\theta}=(\log_{10}(C_eh_\mathrm{1au}^2), \sigma_\Delta, s, q)$ and the known parameters $\bm{X}_k=(M_\star, M_1, M_2, r_1, r_2, j)$. The index $k$ indicates the $k$-th planet pair. 
If a disc structure is specified, the unknown model parameters are $\bm{\theta}=(\log_{10}C_e, \sigma_\Delta)$ for the irradiation disc and $\bm{\theta}=(\log_{10}(C_eh_\mathrm{1au}^2), \sigma_\Delta)$ for the viscous disc, and the known parameters are $\bm{X}_k=(M_\star, M_1, M_2, r_1, r_2, j, h_\mathrm{1au}, s, q)$.
We assume that the prior $p(\bm{\theta}|\bm{X}_k)$ follows a uniform distribution (\tb{prior}). The resonance index $j$ is regarded as one of the known parameters such that we can analyse all pairs at once. 

As mentioned, resonance trapping naturally results from convergent migration: either two planets get trapped into resonance during migration (both inward) with the outer planet migrating faster than the inner one, or the inner planet reaches a migration barrier with the outer planet arriving at a later time. 
Following the discussion in \se{MR-relation}, we assume that the period ratio of a planet pair that is in resonance obeys a log-normal distribution $\log_{10}\Delta\sim\mathcal{N}(\log_{10}\Delta_\mathrm{m/s}(\bm{\theta},\bm{X}_k),\sigma_\Delta^2)$:
\begin{equation}
    \begin{aligned}
        & \quad p_\mathrm{res,m/s}(\Delta|\bm{\theta},\bm{X}_k)d\log_{10}{\Delta} \\
        & = \frac{1}{\sqrt{2\pi\sigma_\Delta^2}} \exp{\left[-\frac{[\log_{10}\Delta-\log_{10}\Delta_\mathrm{m/s}(\bm{\theta},\bm{X}_k)]^2}{2\sigma_\Delta^2}\right]}d\log_{10}{\Delta},
    \end{aligned}
    \label{eq:Gaussian}
\end{equation}
where $\Delta_\mathrm{m/s}$ indicates the resonance commensurability calculated by \eq{commensurability}. The resonance offset $\Delta_\mathrm{m}$ indicates the value calculated for \migrating pairs using \eq{te1_ta1_m}, while $\Delta_\mathrm{s}$ is for \stalling pairs and is calculated using \eq{te1_ta1_s}.

\begin{figure}
    \centering
    \includegraphics[width=\columnwidth]{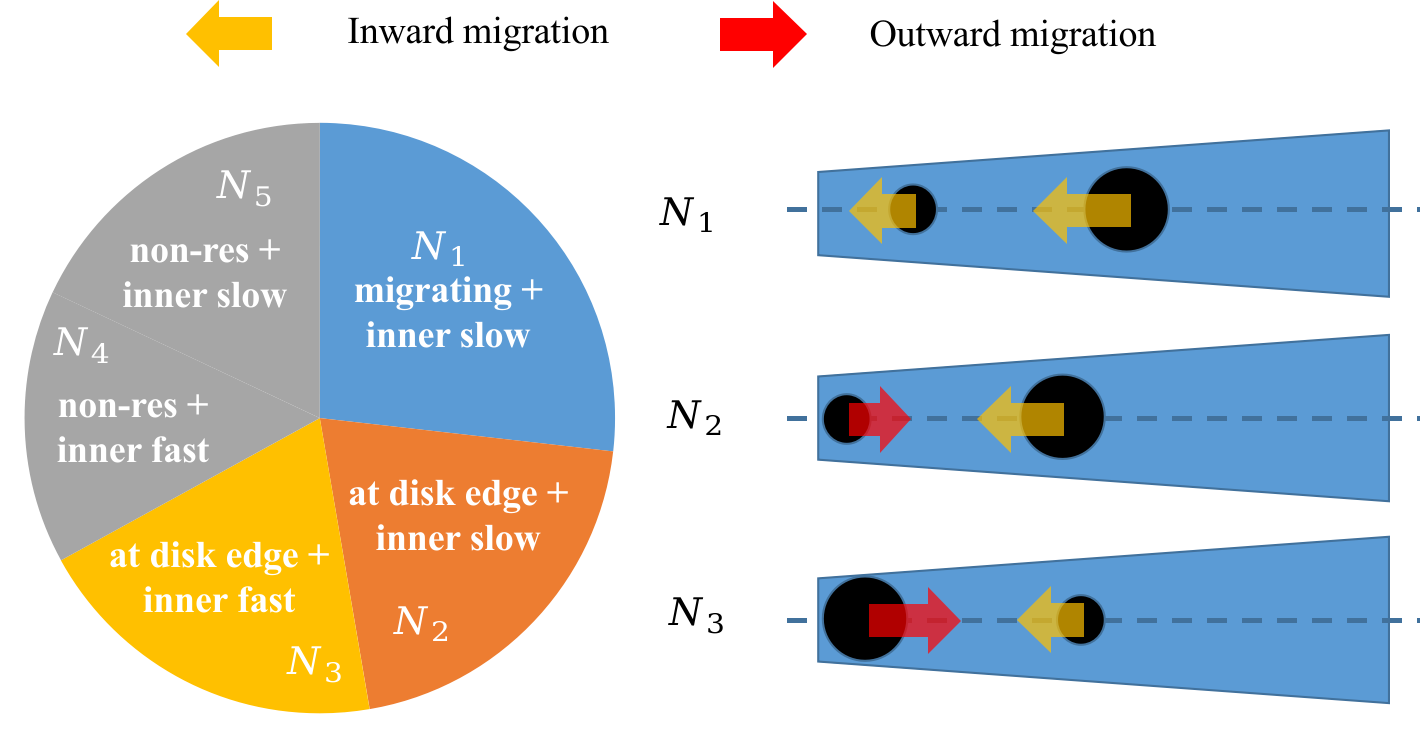}
    \caption{Classification of resonant and non-resonant planet pairs. In the pie chart, 'migrating' means the pairs lock into resonance before reaching a migration barrier, whereas "at disc edge" indicates planets lock into resonance at a planet migration trap. "Inner slow" ("fast") means that the inner planet has a longer (shorter) migration time-scale than its outer planet. $N_\mathrm{res}$ is the number of resonant pairs. $N_1$, $N_2$, and $N_5$ are the numbers of \migrating, \stalling, and non- resonant pairs in the pairs with the inner planets migrating slower than the outer. $N_3$ and $N_4$ are the numbers of \stalling and non- resonant pairs among the pairs with inner planets migrating faster than the outer. }
    \label{fig:model_design}
\end{figure}
In the Type~I migration regime, if the inner planet is more massive (migrates faster), convergent migration can only occur when the inner planet has reached the migration barrier. On the other hand, if the inner planet's migration is slower than the outer, migration is always convergent. In that case, trapping can occur when both planets migrate inward or when the inner planet has reached the migration barrier. 
For ease of the likelihood calculation, we further divide the resonant planet pairs into three categories:
\begin{enumerate}
    \item [Group 1:] The inner planet migrated slower ($\tau_{a_1}>\tau_{a_2}$) and it did not reach a migration barrier ($N_1$);
    \item [Group 2:] The inner planet migrated slower ($\tau_{a_1}>\tau_{a_2}$) and it reached a migration barrier ($N_2$);
    \item [Group 3:] The inner planets migrated faster ($\tau_{a_1}\leq\tau_{a_2}$) and it reached a migration barrier ($N_3$).
\end{enumerate}
$N_1$, $N_2$ and $N_3$ represent the number of resonant planet pairs corresponding to each type of resonance. 
In addition, $N_4$ ($N_5$) represents the number of pairs that are not in resonance with the inner planet migrating faster (slower) than the outer planet.
We provide a sketch to explain the five classes in \fg{model_design}. 

If the inner planets migrate faster ($\tau_{a_1}\leq\tau_{a_2}$), those pairs in resonance must be \stalling pairs. However, if the inner planets migrate slower, they could either be \migrating pairs or \stalling pairs. The $\log_{10}\Delta$ distribution of planets in resonance is therefore (hereafter, $p_{\mathrm{res}}(\Delta)$ represents $p_\mathrm{res}(\Delta_\mathrm{obs}|\bm{\theta},\bm{X}_k)$, etc):
\begin{equation}
\label{eq:pres}
\begin{aligned}
p_{\mathrm{res}}(\Delta)
     & = \left\{\begin{array}{ll}
        p_{\mathrm{res,s}}(\Delta) & \mathrm{if}\, \tau_{a_1}\leq\tau_{a_2}, \\
        \left[p_{\mathrm{res,s}}(\Delta) + p_{\mathrm{res,m}}(\Delta) \right] 
         & \mathrm{if}\, \tau_{a_1}>\tau_{a_2}.
    \end{array}
    \right.
\end{aligned}
\end{equation}


The period ratio of planet pairs that are not in resonance is assumed to follow a uniform distribution:
\begin{equation}
\label{eq:pnres}
    p_{\mathrm{n-res}}(\Delta) d\log_{10}{\Delta} =
        \frac{(\ln{10})\Delta}{\Delta_\mathrm{max}} d\log_{10}{\Delta}
\end{equation}
where $\Delta_\mathrm{max}$ is the value above which a planet pair with $\Delta_\mathrm{obs}$ is not considered in resonance. We use $\Delta_\mathrm{max}=(3j+2)/(3j-1)-(j+1)/j$ which is the distance from the first order $(j+1):j$ resonance to its closest external 3rd order $(3j+2):(3j-1)$ resonance.

Finally, the total log-likelihood is written as:
\begin{equation}
    \begin{aligned}
        \ln\mathcal{L} = \sum_{k=1}^{N_\mathrm{pairs}}\ln\left[p_{\mathrm{res}}(\Delta_{\mathrm{obs},k})+p_{\mathrm{n-res}}(\Delta_{\mathrm{obs},k})\right],
    \end{aligned}
\end{equation}
where $N_\mathrm{pairs}$ is the number of planet pairs in our sample. 


\begin{table}
    \centering
    \begin{tabular}{|c||c|c|c|}
        \hline
        \multirow{2}{3em}{Parameter}    &  \multicolumn{3}{| c |}{Prior}\\
                    &   General & Irradiation & Viscous \\
        \hline
        $\log_{10}(C_eh_\mathrm{1au}^2)$ & $\mathcal{U}(-8,0)$ & - & $\mathcal{U}(-8,0)$ \\
        $\log_{10}C_e$ & - & $\mathcal{U}(-2,2)$ & - \\
        $\sigma_\Delta$ & $\mathcal{U}(0,1)$  & $\mathcal{U}(0,1)$  & $\mathcal{U}(0,1)$  \\
        $s$ & $\mathcal{U}(-5,2.4)$ & $-{15}/{14}$ & $-{3}/{8}$ \\
        $q$ & $\mathcal{U}(-2,2)$ & ${2}/{7}$ & $-{1}/{16}$ \\
        $h_\mathrm{1au}$ & - & 0.0245 & - \\
        \hline
    \end{tabular}
    \caption{Prior bounds or values we take for the disc parameters of the three different models. All priors follow a uniform distribution ($\mathcal{U}$). }
    \label{tab:prior}
\end{table}

\section{Resonance Trapping criterion for the restricted 3-body problem}
\label{sec:new_criterion}
In this section, we improve and numerically verify the two-body resonance trapping criterion. This new trapping condition will be used in \se{upper_limit} to further constrain the statistical results of \se{mcmc_main}.\footnote{{\cite{BatyginPetit2023} have recently presented an analysis with a trapping condition also predicated on the equilibrium phase angle, \eq{sin_phi}, like in this Section. Their findings are consistent with ours.}}

\subsection{Theoretical Derivation}
In the restricted three-body problem, the outer planet is on a fixed circular orbit. The inner planet moves outward and its semi-major axis and eccentricity are damped on time-scales of $\tau_a$ and $\tau_e$, respectively. Lagrange’s planetary equation for the mean motion ($n$) then reads:
\begin{equation}
\label{eq:disturbing0}
    \dot{n}_1=-3j\alpha f_1\mu_2e_1n_1^2\sin{\phi_1}- \frac{3n_1}{2\tau_a}+\frac{p e_1^2n_1}{\tau_e},
\end{equation}
where $\alpha=a_1/a_2$, $\phi_1$, $e_1$ and $f_1$ are the semi-major axis ratio, resonance angle, the eccentricity of the inner planet, and $f_1$ is a numerical factor that depends on the resonance index $j$ \citep{MurrayDermott1999, TerquemPapaloizou2019}. By definition, $p=3$ holds when the eccentricity damping operates at constant angular momentum \citep{TeyssandierTerquem2014}. Lagrange’s planetary equation for eccentricity is:
\begin{equation}
\label{eq:disturbing1}
    \dot{e}_1=-\alpha f_1\mu_2n_1\sin{\phi_1}-\frac{e_1}{\tau_{e_1}},
\end{equation}
When two planets are in resonance, the values of different orbital properties e.g., $e_1$, $\alpha$ and $\phi_1$ librate around their equilibrium values. The equilibrium eccentricity \citep{GoldreichSchlichting2014,TerquemPapaloizou2019} is derived by putting $\dot{e}=\dot{a}=0$ and eliminating $\sin{\phi_1}$ in \eq{disturbing0} and \eq{disturbing1}:
\begin{equation}
\label{eq:RTP_e_eq1}
    e_\mathrm{1,eq}=\sqrt{\frac{\tau_{e_1}}{2(j+1)\tau_{a_1}}}.
\end{equation}
By inserting the equilibrium eccentricity $e_\mathrm{1, eq}$ and $\alpha_\mathrm{eq}\approx[j/(j+1)]^{2/3}$ into \eq{disturbing1}, $\sin\phi_1$ follows:{ 
\begin{equation}
\label{eq:sin_phi}
    \sin \phi_{1,\mathrm{eq}} = -\frac{1}{\alpha f_1\mu_2n_1}\sqrt{\frac{1}{2(j+1)\tau_{e_1}\tau_{a_1}}}
\end{equation}
}
Naturally, its absolute value cannot exceed 1. Otherwise, for $|\sin\phi_1|>1$, no steady state exists and the planets will cross the resonance. 
Combining, $\dot{e}_1=0$, \eq{disturbing1} and \eq{RTP_e_eq1} we can write the resonance trapping condition:
\begin{equation}
\label{eq:new_criterion}
    \tau_{a_1}\tau_{e_1}\geq\frac{1}{2(j+1)(\alpha f_1\mu_2n_1)^2}.
\end{equation}
The classical theory about the resonance trapping criterion is that the time for the planet to migrate across the libration width is shorter than the libration time-scale \citep{OgiharaKobayashi2013, Batygin2015i}. 
For comparison, we also provide the criterion derived from the classical pendulum model \citep{MurrayDermott1999, OgiharaKobayashi2013, HuangOrmel2022i}:
\begin{equation}
\label{eq:classical_criterion}
    \tau_{a_1}\tau_{e_1}\geq\frac{\pi(j+1)}{4(\alpha f_1\mu_2n_1)^2}.
\end{equation}
Compared to \eq{classical_criterion}, the new criterion (\eq{new_criterion}) has the same dependence on planet-to-star mass ratio $\mu_2$ and the orbital frequency $n_1$ but differs regarding the resonance index $j$. 

\subsection{Comparison with simulation}
We compare the new resonance trapping criterion above against the numerical simulation.  The fiducial accelerations accounting for migration and eccentricity of planets in the simulations are expressed by:
\begin{equation}
\label{eq:eom_a}
    \bm{a}_\mathrm{m}=-\frac{\bm{v}}{2\tau_a},
\end{equation}
\begin{equation}
\label{eq:eom_e}
    \bm{a}_\mathrm{e}=-2\frac{(\bm{v}\cdot\bm{r})\bm{r}}{2r^2\tau_e}
\end{equation}
\citep{PapaloizouLarwood2000, CresswellNelson2006, CresswellNelson2008}. We make use of the \texttt{WHfast} integrator of the open-source N-body code \texttt{REBOUND} \citep{Rein2012}. The migration and eccentricity damping on planets are implemented through \texttt{REBOUNDx} \citep{TamayoEtal2020}. 

\begin{figure}
    \centering
    \includegraphics[width=\columnwidth]{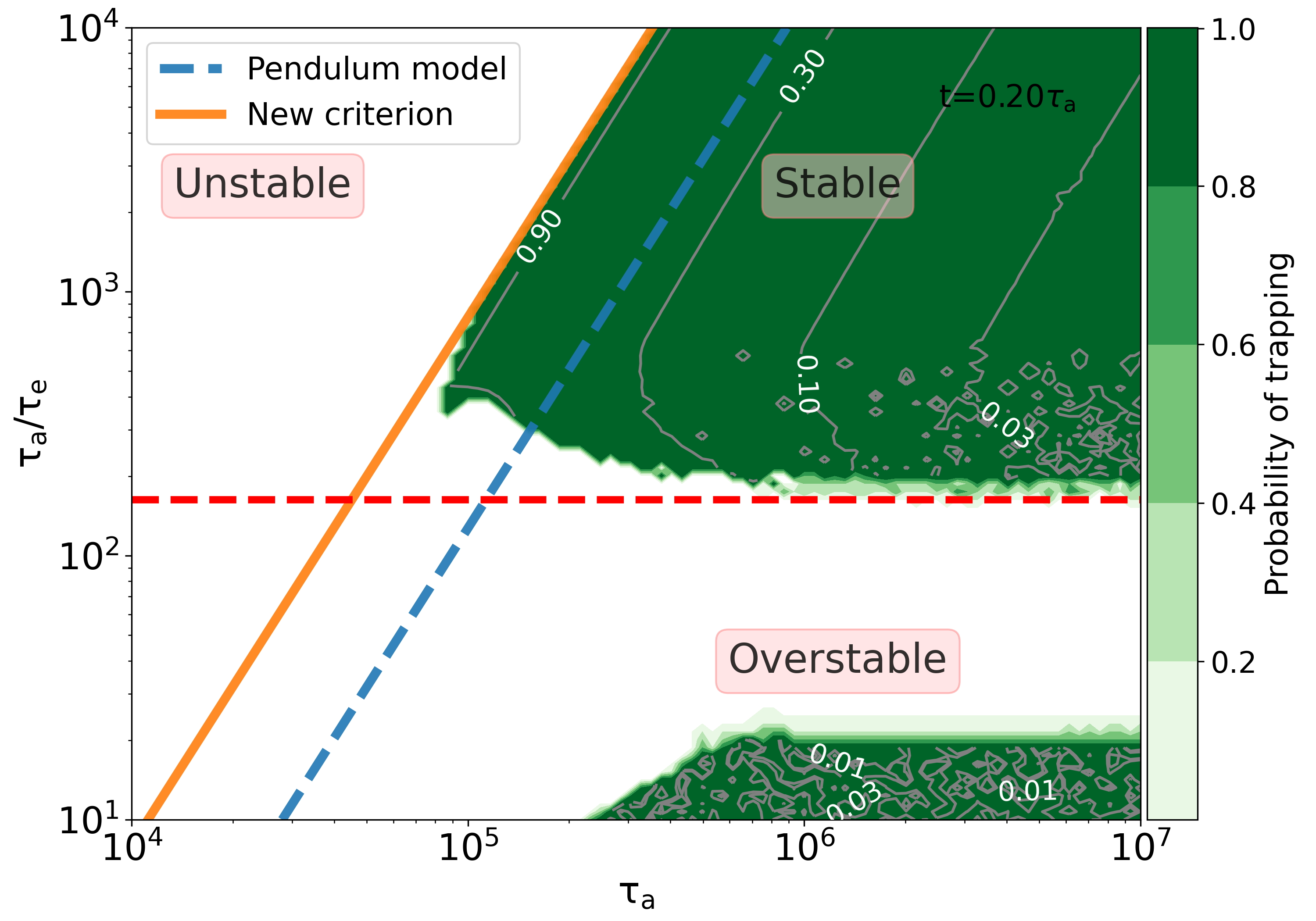}
    \caption{Resonance trapping and crossing for the 2:1 resonance. For each point in the parameter space, we run five simulations with different initial conditions to obtain the probabilistic result of resonance trapping (green shading). 
    The blue line is the resonance trapping criterion derived from the pendulum model (\eq{classical_criterion}) while the orange line is our newly derived criterion (\eq{new_criterion}). {The grey solid lines are the contours of $\sin\phi_1$, with corresponding values labelled in white.} The red dashed line {(\eq{overstable})} corresponds to the transition from stable resonance trapping (above) to overstable resonance (below). Below this line, the simulations progressively evolve into the overstable territory, see \url{https://raw.githubusercontent.com/shuohuangGIT/Infer-migration-history/main/q1.mp4}.}
    \label{fig:q1_20}
\end{figure}

We fix the outer planet on a circular orbit at $[(j+1)/j]^{2/3}$ au. The inner planet starts to migrate outward at $0.8$ au on a time-scale of $\tau_a$. Its eccentricity is damped on a time-scale of $\tau_e$. 
The planet mass is fixed at $10\,M_\oplus$ and the host mass is $1\,M_\odot$. We vary two parameters in the simulation: $\tau_a$ from $10^4$ yr (fast migration) to $10^7$ yr (slow migration), and $\tau_a/\tau_e$ ranging from $10^1$ (inefficient eccentricity damping) to $10^4$ (efficient eccentricity damping). {Each parameter is sampled by 100 grid points evenly distributed in log-space.} In order to capture the probabilistic behavior of resonance trapping, we run five simulations for each point in the $\tau_a/\tau_e$--$\tau_a$ parameter space, where we evenly sample the initial longitude of the inner planet.

We conduct simulations for $j=1$ (2:1 resonance) and $j=2$ (3:2 resonance). We run the simulation until $t=\tau_a$, but we take a snapshot at $t=0.2\tau_{a}$. If the period ratio $P_2/P_1$ decreases below $(j+1)/j$, we classify the simulation outcome as a resonance crossing. 
The results are shown in \fg{q1_20} and \fg{q2_20} for $j=1$ and $j=2$ respectively. Resonance crossing cases are in white and resonance trapping cases are in green. The red dashed line indicates the boundary below which the trapping solution becomes overstable,
\begin{equation}
\label{eq:overstable}
    \left( \frac{\tau_a}{\tau_e} \right)_\mathrm{overstable} = \frac{1}{8(j+1)}\left(-\frac{3 j^2}{\alpha f_1 \mu_2}\right)^{2/3}
\end{equation}
\citep{GoldreichSchlichting2014}, which evaluates to
$\tau_{a}/\tau_{e}\approx${170 for $\mu_2=3\times10^{-5}$, both for $j=1$ and 2}. Above this line, all systems are either trapped in resonance or not. The top-right corner indicates the parameter space where the two planets both get captured and permanently stay in resonance and the top-left indicates resonance crossing. Below this line, some systems are still evolving and resonance trapping is only temporary. 

\begin{figure}
    \centering
    \includegraphics[width=\columnwidth]{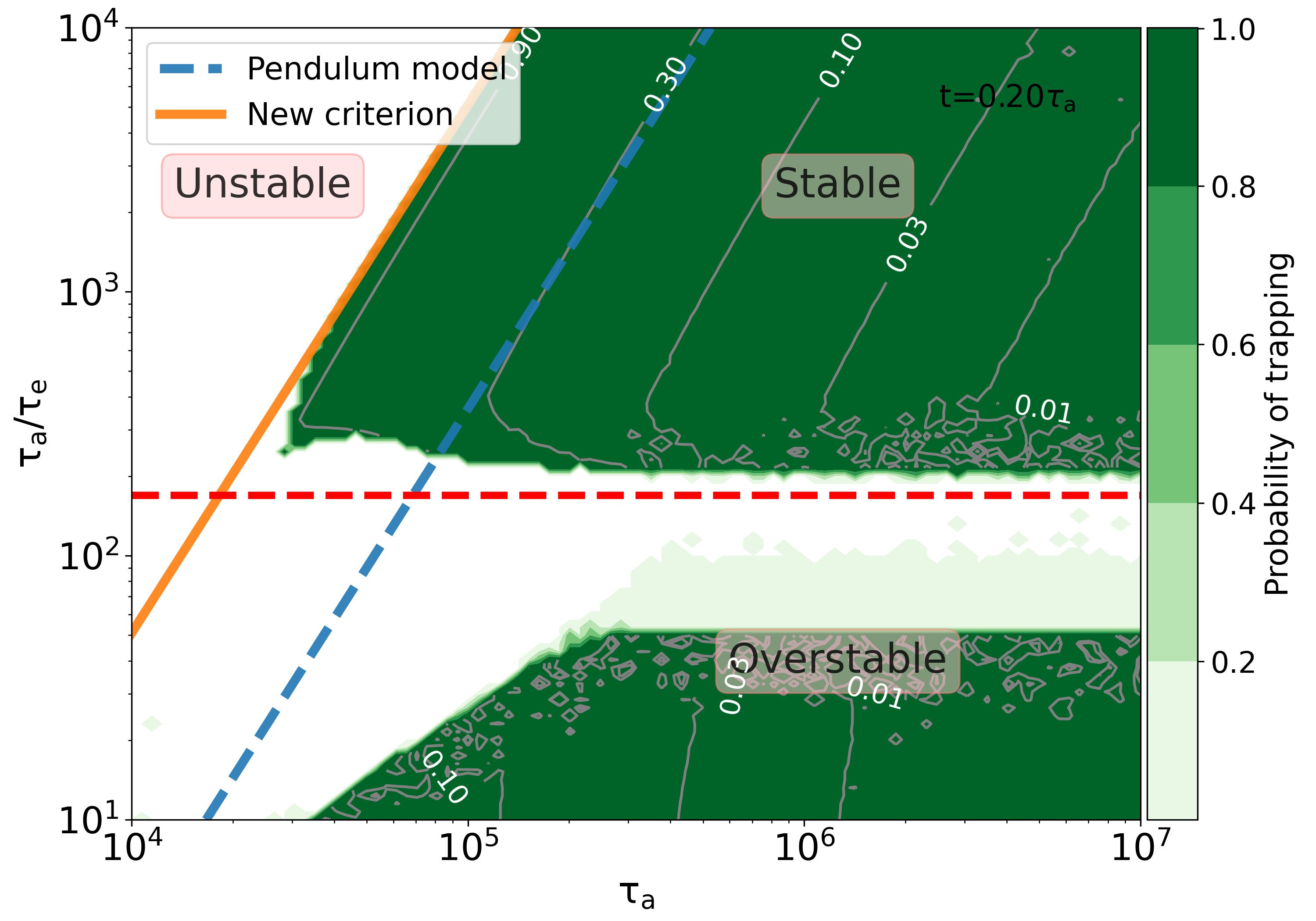}
    \caption{Same as \fg{q1_20}, but for the 3:2 resonance. The video can be downloaded on Github: \url{https://raw.githubusercontent.com/shuohuangGIT/Infer-migration-history/main/q2.mp4}. }
    \label{fig:q2_20}
\end{figure}

We indicate the trapping criterion derived from the pendulum model in blue and the improved trapping criterion (\eq{new_criterion}) with the orange line. From Fig. \ref{fig:q1_20} and \ref{fig:q2_20} it is clear that the pendulum model criterion for resonance trapping (blue line) fails to quantitatively match the numerical simulations. Our new criterion (orange line), however, fits the simulations perfectly.
{{The equilibrium value of $\sin\phi_\mathrm{1}$ for the simulation snapshots is calculated by averaging its value over a time span of $0.1\tau_a$ before and after the snapshot time, e.g., $0.1-0.3\tau_a$ for the snapshot at $t=0.2\tau_a$.} The values of $\sin\phi_\mathrm{1,eq}$ is indicated by grey solid lines (contour) in Fig. \ref{fig:q1_20} and \ref{fig:q2_20}. $\sin\phi_\mathrm{1,eq}$ increases as getting closer to the orange line, which is also expected by \eq{sin_phi}.}
The picture of resonance trapping/crossing over the entire parameter space of the migration time-scale and eccentricity damping time-scale has now been clarified. Migration plays a role in exciting the planet's eccentricity, while eccentricity damping reduces it. On one hand, if a planet pair is in resonance, the eccentricity damping balances its excitation and finally librates near the equilibrium value. If the migration speed is so fast that there is no steady state solution for the resonance angle $\phi_\mathrm{1, eq}$ (\eq{new_criterion}), the resonance is crossed. Otherwise, resonance trapping is ensured. On the other hand, if eccentricity is excited to be high enough, planets can be captured into resonance, but only temporarily, because of the continuous increase of the resonance libration amplitude \citep[overstability, cf.][]{GoldreichSchlichting2014}. {Although not evident from the figures presented, all simulations located in the large green corner above the red dashed line exhibit permanent libration of $\sin\phi_\mathrm{1}$, indicating that the planets are captured in resonance permanently. The amplitude of $\sin\phi_\mathrm{1}$ during libration increases as the ratio $\tau_a/\tau_e$ decreases, and approaches the red dashed line denoting overstability in Fig. \ref{fig:q1_20} and \ref{fig:q2_20}. The green region situated below the red dashed line represents simulations in which planets are temporarily captured in resonance, with their amplitude of $\sin\phi_\mathrm{1}$ increasing over time and circulating at the end of the simulation.} In conclusion, both efficient eccentricity damping ($\tau_a/\tau_e$ is high) and slow migration are required for permanent resonance trapping.

\begin{figure}
    \centering
    \includegraphics[width=\columnwidth]{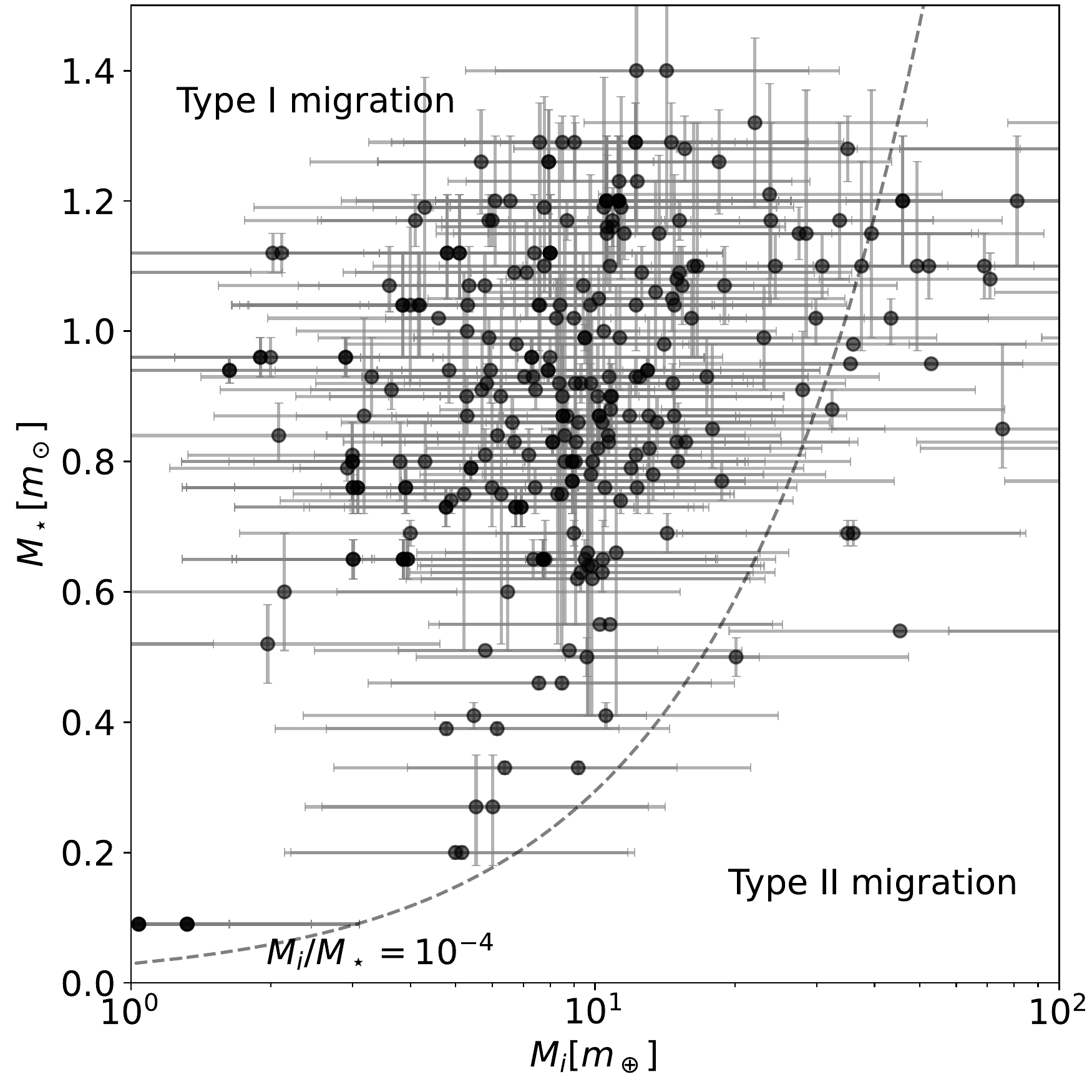}
    \caption{Planet mass versus stellar mass in the sample. Planet pairs with period ratios larger than $(j+1)/j$ and smaller than $(3j+2)/(3j-1)$ are chosen. We take $j$ equal to $1$, $2$, $3$ and $4$. Planet masses and their host masses are indicated by black dots, with 1$\sigma$ error bar. If the planet mass is inferred from its radius through the M-R relation, its uncertainty of the masses is assigned to be $\sigma_\mathrm{m}=0.374$ dex. The grey dashed line indicates the position where the planet-to-star mass ratio equals {$\mu=10^{-4}$}. We assume planets to the left of this line to follow Type I migration and otherwise Type II. Only Type~I migrating planets are included in our analysis. }
    \label{fig:mass_dist}
\end{figure}
\begin{figure*}
    \centering
    \includegraphics[width=0.8\textwidth]{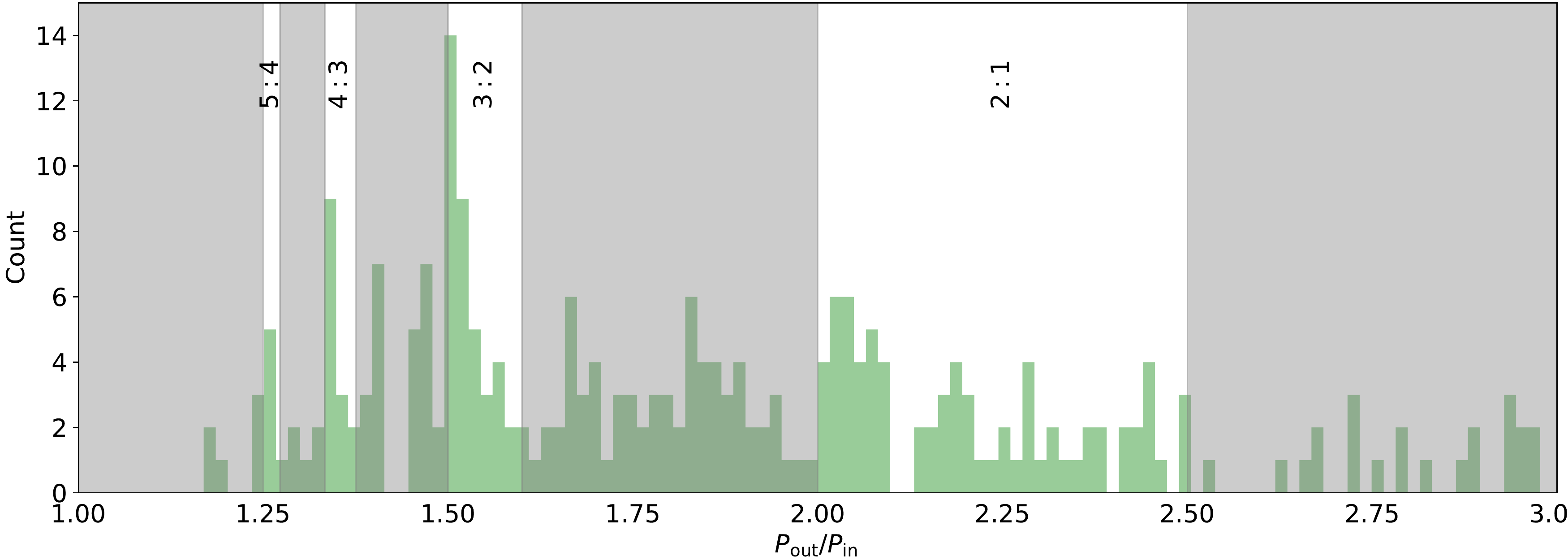}
    \caption{Distribution of the period ratio of all observed planet pairs in our sample (green histograms). We highlight the four windows in the vicinity of four different first-order resonances: 2:1, 3:2, 4:3 and 5:4. The left boundary of each window is $(j+1)/j$ and the right boundary is $(3j+2)/(3j-1)$. The total number of pairs is 371 in the green histogram and 128 in the windows. }
    \label{fig:per_dist_sample}
\end{figure*}

\section{MCMC analysis}
\label{sec:mcmc_analysis}
In this section, we first discuss how we select our sample in \se{sample}. Then we conduct an MCMC fitting to constrain the model parameters. 

\subsection{Sample selection}
\label{sec:sample}
Our sample selection and all analysis are based on the NASA Exoplanet Archive. Our attention is drawn to planets detected through transits and TTV. As shown in \se{resonance}, we need planets' masses to calculate the equilibrium eccentricities and period ratios in resonance. If only the radius is available, the M-R relationship described in \se{MR-relation} is used to calculate the masses of those transiting planets. Additionally, the semi-major axes and stellar masses are extracted. 

We do not consider short-period planets in our sample. We take this simple step to reduce the effects of both photo-evaporation and stellar tides on the M-R relationship. 
First, photo-evaporation can alter the M-R relationship for planets with low-density atmospheres on time-scale of 1 Gyr \citep{FultonPetigura2018}. It is believed to have triggered the so-called `radius gap' \citep{FultonEtal2017}. When planets get closer to their host stars, this effect is more obvious \citep{FultonPetigura2018}. 
Second, stellar tides alter close-in planets' orbital properties \citep{LithwickWu2012, BatyginMorbidelli2013, CharalambousEtal2018, PapaloizouEtal2018} and blur the information of the planets inherited from their protoplanet disc. Both mechanism are very sensitive to planets' semi-major axes. 
Excluding the planets with a cutoff period shorter than 5 days, though crude, can suppress the interference from stellar tides \citep{ChoksiChiang2020} and photo-evaporation \citep{FultonPetigura2018} on our sample. 

We display the planet mass versus their host mass in \fg{mass_dist}. The average planet mass is $10M_\oplus$ and the average stellar mass is $1\,M_\odot$. The planets' and stars' mass uncertainties are indicated by error bars. If the planet mass is inferred from the M-R relation, the log-normal standard deviation is then $\sigma_m=0.374$ (\eq{MRfitting}). The migration speed of low-mass planets in the proto-planetary disc scales linearly with planet mass, as dictated by the Type~I migration limit.
As planets become massive enough to perturb their surrounding disc, their migration gradually switches to Type II \citep{KanagawaEtal2018, PichierriEtal2022}. We set the boundary between two types of migration as {$\mu_\mathrm{trans}=10^{-4}$}. Since our interests focus on Type~I migration only planets with {$\mu<\mu_\mathrm{trans}$} are included in our sample. 

Finally, the sample size is reduced to 371 and the period ratios for all planet pairs are given by \fg{per_dist_sample}. {The selected planets come from systems with two and more planets, including those with resonance chains. } 

\begin{figure}
    \centering
    \includegraphics[width=\columnwidth]{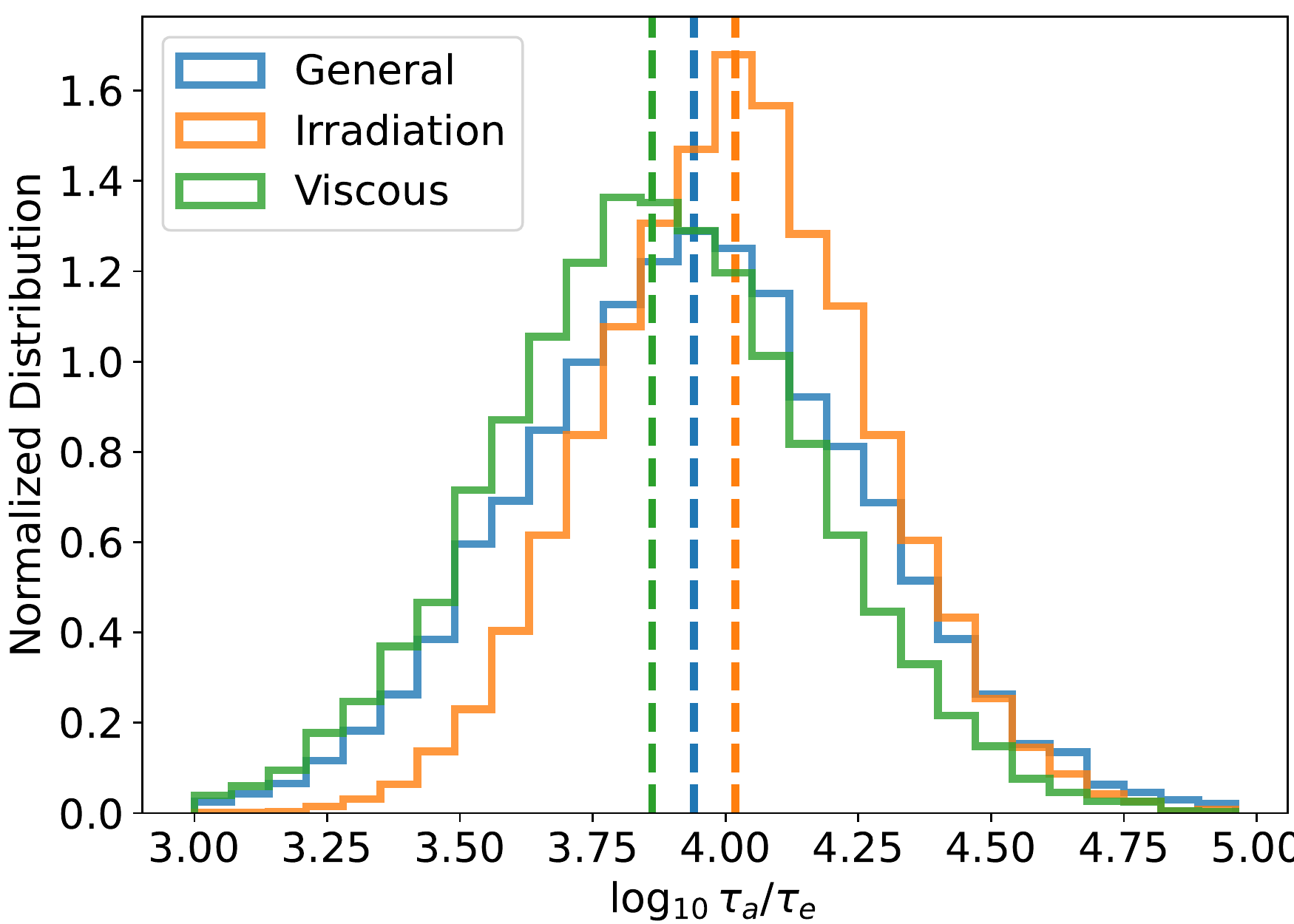}
     \caption{Distribution of $\log_{10}\tau_a/\tau_e$, the semi-major axis-to-eccentricity damping time-scale at the location of the inner planets averaged over all planet pairs, in General (blue), Irradiation (orange) and Viscous (green) model. We calculate this quantity based on the posterior distributions of the parameters in each model. Dashed lines indicate their median values. }
    \label{fig:ta_te_dist}
\end{figure}

To calculate the resonance offset, the resonance number $j$ is required. There are excesses of systems ('peaks') just wide of the integer period ratios, which is suggestive of resonances. We only consider first-order resonances: 2:1, 3:2, 4:3, and 5:4. We assume that planets with period ratios slightly larger than the integer ratios are potentially in resonance until the period ratio "hits" the {third}-order resonance, because the resonant interaction is weaker as they become further from exact commensurability. Planet pairs with period ratios larger than $(j+1)/j$ but smaller than $(3j+2)/(3j-1)$ are possibly in $(j+1)/j$ resonance. Here, $(3j+2)/(3j-1)$ is the location of the closest third-order resonance. {The selection of a period ratio limit for identifying planets in 2:1 resonance may seem arbitrary, given that the period ratio can extend up to 2.5, where planet pairs are unlikely to be in resonance. However, a slightly smaller window for the 2:1 resonance would not affect our conclusions. Nonetheless, this choice is useful for identifying planets in 3:2, 4:3, and higher $j$ first-order resonances because planets located near these resonance locations are close to nearby higher-order resonances, and may therefore be more easily perturbed.} The satisfied period ratio windows are highlighted in \fg{per_dist} (top panel) and the four lower panels zoom in on these four windows, for $j=$1, 2, 3, 4, where we instead show the distribution of the offset from resonance, $\Delta$. Planets out of the windows may still be in first-order resonance, but their fraction must be very low and it is not covered by our analysis. We ignore other first-order resonances and all higher-order resonances.

\subsection{Implication on planet-disc interaction from MCMC}
\label{sec:mcmc_main}
We use \texttt{emcee} \citep{Foreman-MackeyEtal2013} to perform the MCMC analysis. We implement three different models: 
\begin{enumerate}
    \item General model. The disc structure is not specified and the MCMC is used to fit $\log_{10}(C_eh_\mathrm{1au}^2)$, $\sigma_\Delta$, $s$, $q$. 
    \item Irradiation model. Stellar irradiation is assumed to be the main heating source and the MCMC is used to fit $\log_{10}C_e$ and $\sigma_\Delta$.
    \item Viscous model. Viscosity-driven accretion is assumed to be the main heating source and the MCMC is used to fit $\log_{10}(C_eh_\mathrm{1au}^2)$, $\sigma_\Delta$.
\end{enumerate}
The prior distribution of parameters $\log_{10}(C_eh_\mathrm{1au}^2)$, $\log_{10}C_e$, $\sigma_\Delta$, $s$, $q$ and the values we take for $s$, $q$ and $h_\mathrm{1au}$ for the three different models are shown in \tb{prior}. 
{The convergence of MCMC chains are checked. We make use of the criterion that MCMC converges if the autocorrection time is shorter than 1/50 times its chain length. We checked that our results all satisfied the convergence criterion.  }

For the General model, we examine whether our method is capable to retrieve all the parameters in \Ap{model_test}. It turns out that almost all parameters are degenerate. Therefore, the fitted values for $\{\log_{10}(C_eh_\mathrm{1au}^2), \sigma_\Delta, s, q\}$ may not be reliable (\Ap{model2_test_general}). The result of the General model is shown and analysed in \Ap{mcmc_general}. We also calculate the quantity $\log_{10}\tau_a/\tau_e$, the semi-major axis-to-eccentricity damping time-scale, at the location of the inner planets averaged over all planet pairs. This quantity, however, shows to be independent of the other parameters and can be reproduced within the 1$\sigma$ error bar (\Ap{model2_test_general}). 

We calculate $\log_{10}\tau_a/\tau_e$ in all three models, and their distributions are shown in \fg{ta_te_dist}. Two key points can be made. First, different disc structures result in nearly identical distributions. The parameter $\log_{10}\tau_a/\tau_e$ is not sensitive to the assumed disc structure.
Second, the value of $\log_{10}{\tau_a}/{\tau_e}$ -- peaking at 4 and almost always larger than 3 -- is high. The {high semi-major axis-to-eccentricity damping time-scale ratio} indicates that temporary capture (overstable libration) did not operate for the planets in our sample, which would require  $\tau_a/\tau_e\approx${170} \citep{GoldreichSchlichting2014} in \eq{overstable}. 

By specifying the disc structure -- the Irradiation or Viscous model -- the parameters can be successfully retrieved within 1$\sigma$ error bar (\Ap{model2_test}). 
We show the fit result of $\{\log_{10}C_e,\ \sigma_\Delta\}$ and $\{\log_{10}(C_eh_\mathrm{1au}^2)$, $\sigma_\Delta\} $ 
for the Irradiation and Viscous model in \fg{mcmc_irradiation} and \fg{mcmc_viscous}, respectively. The python package \texttt{corner.py} \citep{corner} is used to generate the plots of the posterior distributions. 

The Irradiation model (\fg{mcmc_irradiation}) fits $\log_{10}C_e=0.24^{+0.25}_{-0.25}$. For the viscous model, the disc aspect ratio is sensitive to the stellar accretion rate and disc opacity. Therefore, we fit the combination $C_e h_\mathrm{1au}^2$, and $\log_{10}(C_eh_\mathrm{1au}^2) =-3.60^{+0.30}_{-0.29}$ (\fg{mcmc_viscous}). If we take $h_\mathrm{1au}=0.0245$, same value as the irradiation model, then $\log_{10}C_e=-0.38^{+0.30}_{-0.29}$. Increasing $h_\mathrm{1au}$ (or $L_\star$ in \eq{h_g_irr}) would result in a smaller value of $C_e$ ($C_e\propto h_\mathrm{1au}^{-2}$). Theoretically, \cite{TanakaEtal2002} and \cite{TanakaWard2004} from the first principle calculate that $\log_{10}C_e=0$ for locally isothermal discs. The fitting outcomes from both models are consistent with it.

Additionally, the fitted values for $\sigma_\Delta$ are $0.65^{+0.10}_{-0.08}$ and $0.79^{+0.10}_{0.09}$ for the Irradiation and Viscous model, respectively. Their values are twice that of the mass dispersion. Indeed, we expect that the fitted $\sigma_\Delta$ is of the same magnitude as $\sigma_\mathrm{m}$ (\se{MR-relation}). However, $\sigma_\Delta$ is fitted to be slightly larger than our expectation. This could be an implication of turbulent discs \citep{ReinPapaloizou2009,GoldbergBatygin2022} and/or post-disc perturbations \citep[e.g.][]{LithwickWu2012,ChatterjeeTan2014,StockEtal2020}. We further run an MCMC fitting fixing $\sigma_\Delta$ to 0.374, the resulting posterior distribution of $\log_{10}C_e$ or $\log_{10}(C_eh_\mathrm{1au}^2)$ are not significantly different from what we present here. It gives $\log_{10}(C_eh_\mathrm{{1au}}^2)=-3.95^{+0.29}_{-0.27}$ for the Viscous model and $\log_{10}C_e=0.44^{+0.19}_{-0.20}$ for the Irradiation model.

\begin{figure}
    \centering
    \includegraphics[width=0.9\columnwidth]{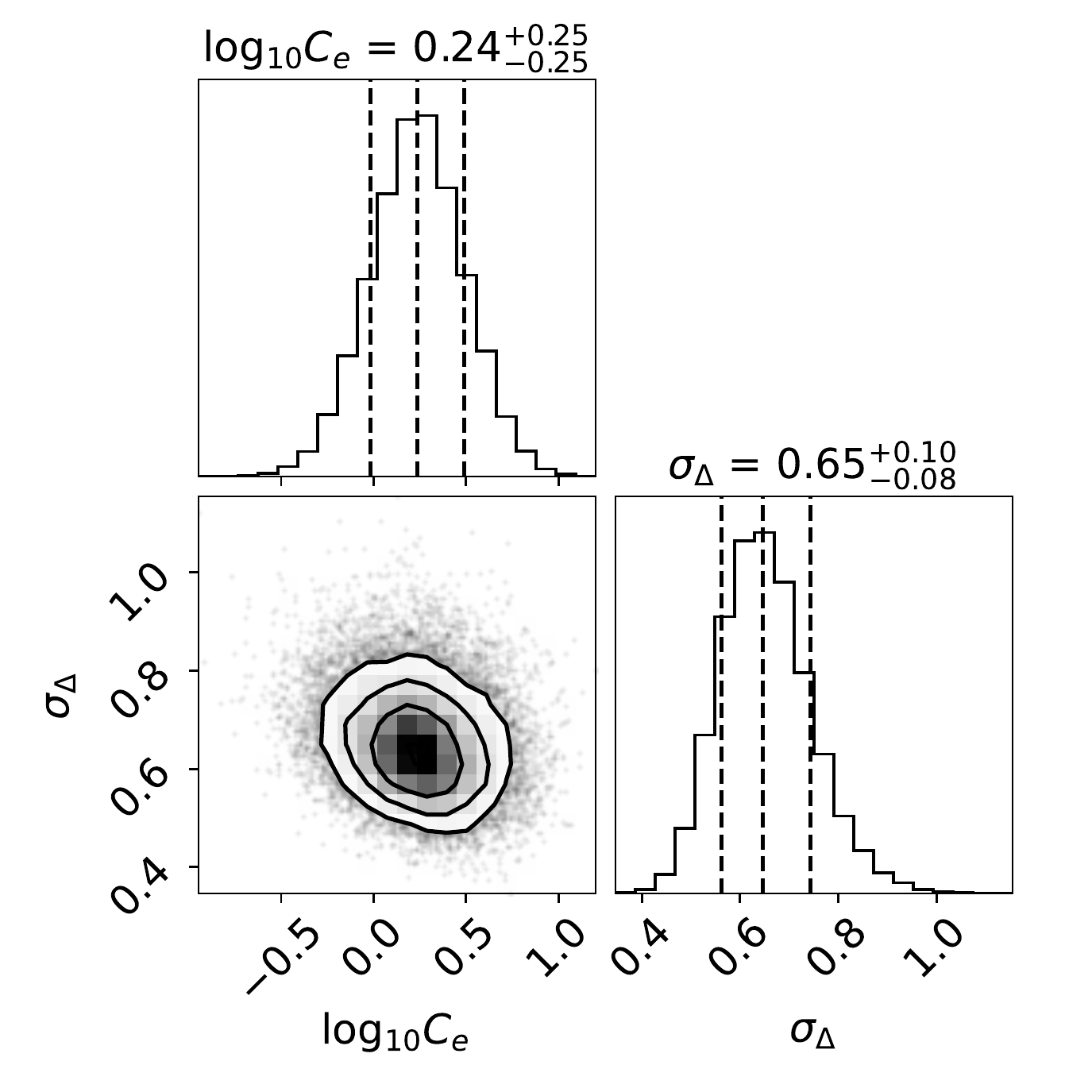}
    \caption{
    Corner plot of variables in the MCMC analysis ($\log_{10}C_e$ and $\sigma_\Delta$) with 1$\sigma$, 2$\sigma$ and 3$\sigma$ confidence contours, for the Irradiation model, which fixes the surface density power law index $s=-15/14$ and the disc aspect ratio index $q=2/7$. {The 1$\sigma$ uncertainty is labelled on the top of each column and indicated by left and right dashed lines. The middle dashed lines indicate their median values.}}
    \label{fig:mcmc_irradiation}
\end{figure}

In summary, our MCMC model shows that eccentricity damping is highly effective ($\log_{10}\tau_a/\tau_e\approx 4$), making resonant over-stability unlikely. {The observed period ratio excess of planets is consistent with predictions by \citet{TanakaEtal2002} and \citet{TanakaWard2004} ($C_e\approx1$), irrespective of whether the disk structure is dominated by irradiation or viscous heating. However, the aspect ratio of a viscous inner disc depends on the disc opacity and stellar accretion rate \citep[e.g.][]{LiuEtal2019}, which limits our ability to constrain $C_e$.}

\section{Implications for planet formation}
\label{sec:implication}
In this section, we adopt the fit result from the Irradiation model and further study the implications of resulting resonant planets statistically. \cite{RamosEtal2017} and \cite{CharalambousEtal2022} use similar prescription for their disc structure. The reason why we choose the Irradiation model is the following. Even though resonance trapping can happen much earlier, planets' period ratios (offsets) are more evolved at the end of the disc lifetime when the migration and eccentricity damping time-scales are longer than the disc dispersal time-scale. The disc structure at this stage mostly determines what the corresponding mature planet system looks like. Because planet formation consumes solids and solids drift inward rapidly due to gas drag \citep{Weidenschilling1977, AndrewsEtal2012}, the disc at this point becomes optically thin, rendering stellar irradiation the main heating source. The transition disc LkCa15 is arguably an example that low-mass planets can carve a large dust cavity \citep{LeemkerEtal2022}. Therefore, the Irradiation model is more applicable to transition discs.

The best-fitting resonance offset distribution is plotted in \fg{per_dist} (blue lines), with upper and lower 3$\sigma$ uncertainty (blue shaded regions). We assume that the offset of non-resonant pairs follows a uniform distribution, which is also indicated in \fg{per_dist} (grey dashed lines). The MCMC fits the observed distribution better than the uniform-only model because it fits more planets with small $\Delta$ and fewer planets with large $\Delta$, just as observed. 
{The complete sample with 128 pairs are fitted simultaneously. However, we use four panels to display the four near-resonance planets because $\Delta$ depends on the resonance number $j$ in a complex form. We cannot present one distribution of $\Delta$ to represent all 128 pairs while keeping the shape of log-normal profile.} 

For each planet pair, we calculate its probability of being in resonance ($P_{\mathrm{res},k}$, in \se{resonant_fraction}). The total number of resonant pairs and planets' mean eccentricities in our sample is then obtained. The properties of their birthplace -- the natal proto-planet disc -- are then inferred, e.g., the upper limit of the surface density (\se{upper_limit}) and the location of the migration barrier (\se{barrier}).  

\begin{figure}
    \centering
    \includegraphics[width=0.9\columnwidth]{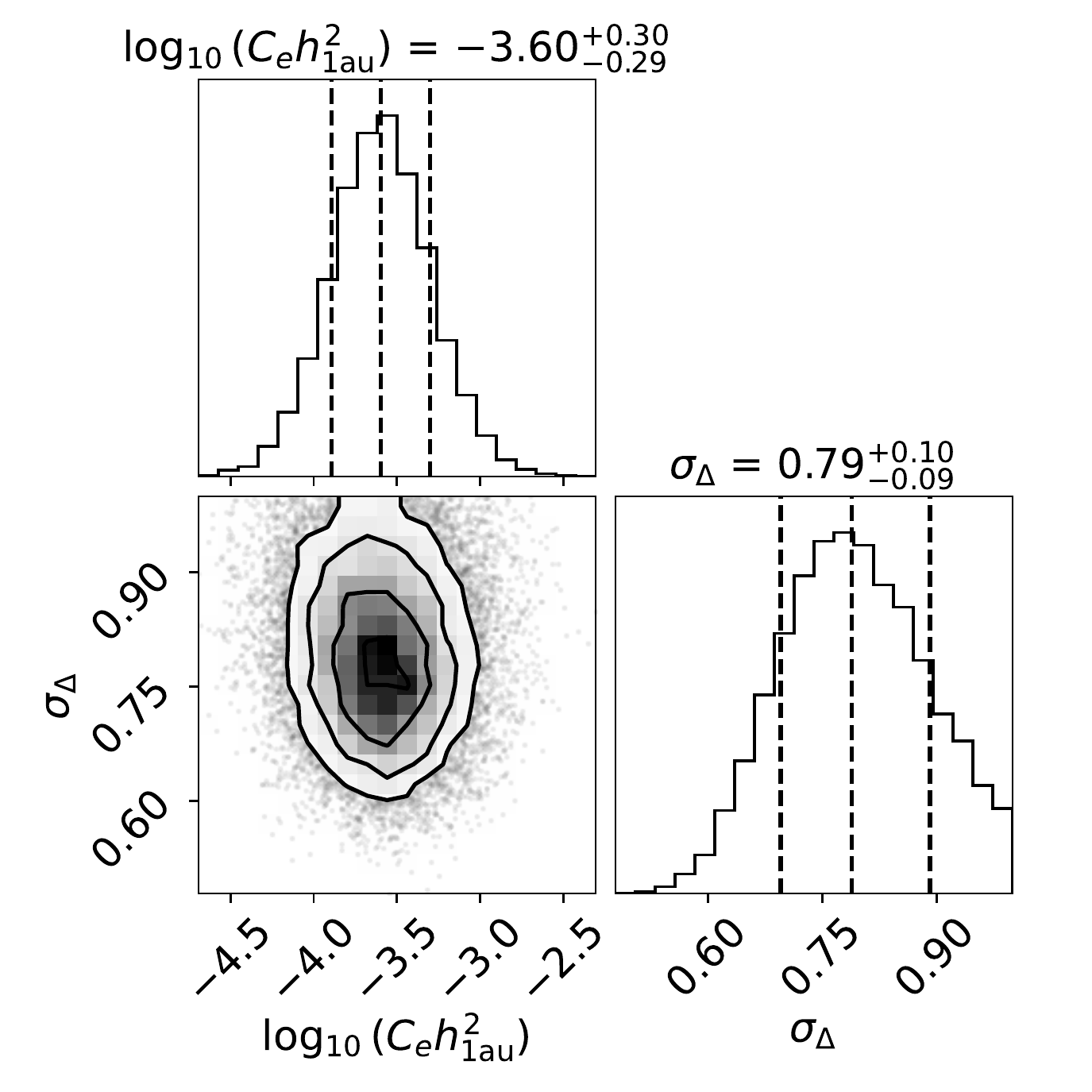}
    \caption{
    Similar to \fg{mcmc_irradiation}, but for the Viscous model. The surface density power law index is fixed at $s=-3/8$ and the disc aspect ratio index at $q=-1/16$. }
    \label{fig:mcmc_viscous}
\end{figure}

\begin{figure*}
    \centering
    \includegraphics[width=0.9\textwidth]{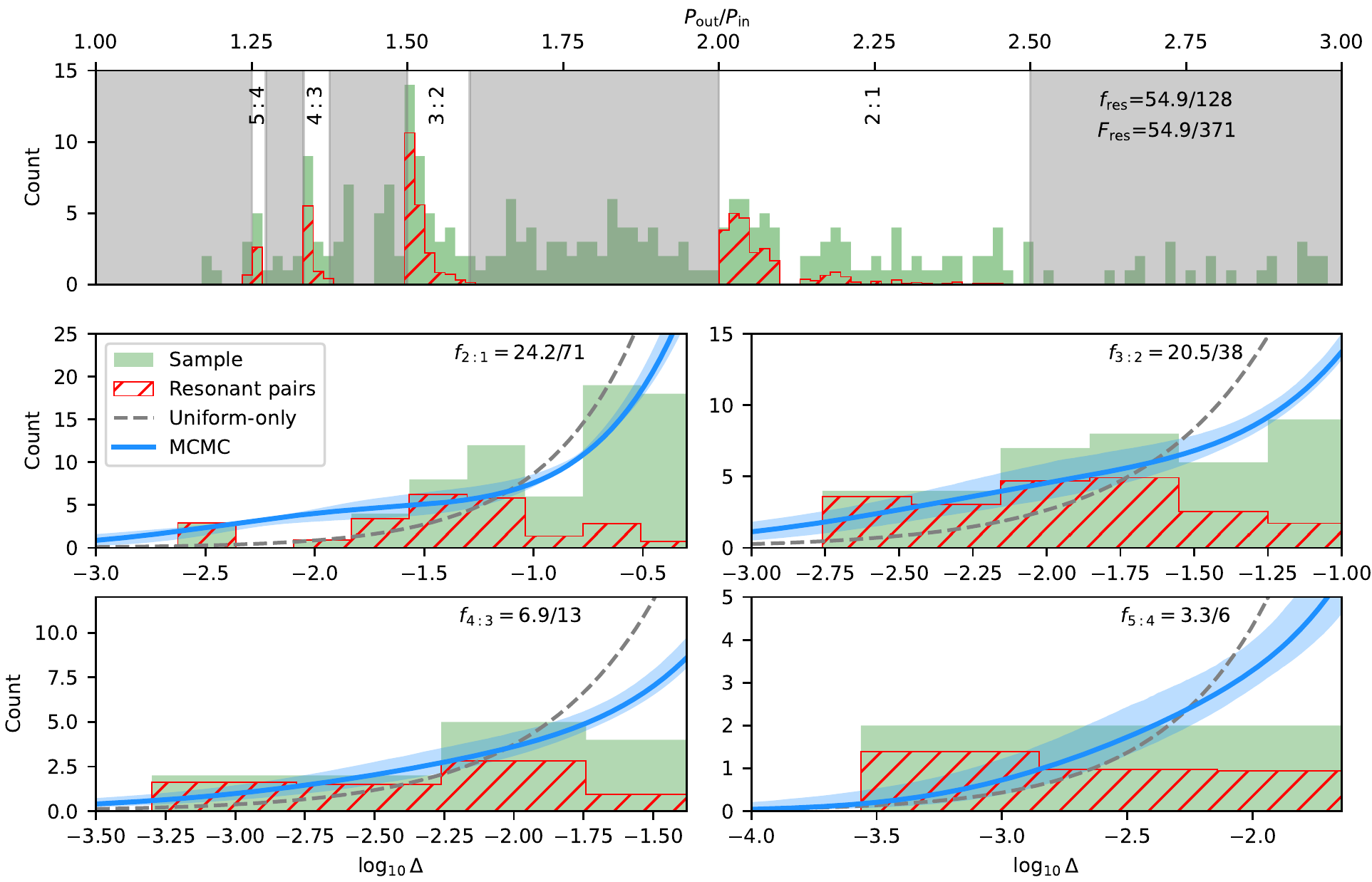}
    \caption{Distribution of period ratio (top panel, similar to \fg{per_dist_sample}) and resonance offset $\Delta$ (bottom panel) for the inferred resonant planet pairs (red hatches) versus all observed planet pairs in our sample (green histogram).
    The lower panels display the distribution of period ratio offset $\Delta$ within the highlighted four windows in the top panel. Grey dashed lines show the probability distribution of all planet pairs assuming they follow a uniform distribution (not in resonance). 
    We compare it with the best MCMC fit result (blue curve, after bin size correction) assuming the Irradiated disc model. The shaded blue region indicates the 3$\sigma$ uncertainty. We also label the fraction of inferred resonant planets, which is the ratio of the area between the red hatches and the green histograms (see \se{resonant_fraction} for details).}
    \label{fig:per_dist}
\end{figure*}

\subsection{Fraction of resonant pairs}
\label{sec:resonant_fraction}

Given the fitted value for $C_e$ and $\sigma_\Delta$, we can calculate the probability of each planet pair in resonance via two characteristic probability distribution functions (\eq{pres} and \eq{pnres}): $P_{\mathrm{res},k}={p_{\mathrm{res}}(\Delta_{\mathrm{obs},k})}/[{p_{\mathrm{res}}(\Delta_{\mathrm{obs},k})+p_{\mathrm{n-res}}(\Delta_{\mathrm{obs},k})}]$. In \fg{per_dist}, we plot the histogram of period ratio and resonance offset distribution of planet pairs weighted by $P_{\mathrm{res},k}$ (red hatches). It indicates the period ratio and resonance offset distribution of resonant pairs. The period ratio of resonant pairs peaks just wide of integer ratios, and, as the period ratio further increases, resonant pairs vanish. Not all pairs with period ratios close to integer ratio are in resonance. 

The total number of resonant pairs is the summation of resonant probability over all planet pairs: 
\begin{equation}
    N_\mathrm{res} = \sum_{k=1}^{N_\mathrm{pairs}} P_{\mathrm{res},k}.
\end{equation}
We label the average fraction of resonant pairs, $f_\mathrm{res}={N_\mathrm{res}}/{N_\mathrm{pairs}}$, on the top right of four lower panels in \fg{per_dist}. $N_\mathrm{pairs}$ is the number of pairs in the narrow period ratio windows. 

We also calculate the fraction of all {resonant pairs} among all pairs in our sample: $F_\mathrm{res}=N_\mathrm{res}/371=14.8^{+0.5}_{-0.7}\%$. The distribution is shown in \fg{res_frac} left panel. It is consistent with the crude estimation made by \citet{WangJi2014} (10\%-20\%). This number ignores higher-order resonances and first-order resonances with resonance numbers larger than 4 (5:4). The resonant fraction could therefore be higher. 

We split the resonant planets into three groups, each group has a number of resonant pairs $N_1$, $N_2$, $N_3$, respectively, and $N_\mathrm{res}=N_1+N_2+N_3$ (see \se{MCMC_model} for detail and \fg{model_design} for a sketch). Three different resonant fractions are calculated: 
\begin{itemize}
    \item $f_\mathrm{res}\mathrm{(Mig|out)}={N_1}/{(N_1+N_2+N_5)}$: fraction of \migrating pairs among the pairs where the inner planets migrate slower than the outer; 
    \item $f_\mathrm{res}\mathrm{(Brk|out)}={N_2}/{(N_1+N_2+N_5)}$: fraction of \stalling pairs among the pairs where the inner planets migrate slower than the outer;
    \item $f_\mathrm{res}\mathrm{(Brk|inn)}={N_3}/{(N_3+N_4)}$: fraction of \stalling pairs among the pairs where the inner planets migrate faster than the outer.
\end{itemize}
This classification allows us to compare the fraction of resonant pairs under different physical conditions. The distributions of the three fractions are shown in \fg{res_frac}, right panel. It shows that $f_\mathrm{res}\mathrm{(Mig|out)} \approx 0.27$. This implies that in some pairs the ambient gas disc disperses before the pair reaches a migration barrier, i.e., either planet migration is slow or the disc disperses rapidly following planet migration and formation. The gas-poor formation scenario for sub-Neptunes \citep{DawsonEtal2015, ChoksiChiang2020} would be consistent with this picture. However, a still larger fraction of resonant pairs, $f_\mathrm{res}\mathrm{(Brk|out)} \approx 0.23$, and $f_\mathrm{res}\mathrm{(Brk|inn)} \approx 0.41$ reach their migration barriers. It implies that gas-poor formation applies to some of the observed systems, but not all of them.  Interestingly, $f_\mathrm{res}\mathrm{(Brk|out)}$ is smaller than $f_\mathrm{res}\mathrm{(Brk|inn)}$, i.e., the inner planets tend to be more massive than the outers. A possible explanation would be that the inner edge of the disc -- the location of the pressure maximum -- would also be the place where pebbles accumulate. Pebble accretion at such locations can be very efficient \citep{ChatterjeeTan2014, JiangOrmel2023}.

We also plot the eccentricity distribution  for the planets in resonant pairs, weighted by their probability of being in resonance ($P_{\mathrm{res},k}$), in \fg{ecc_dist}. Generally, the outer planets have smaller eccentricities consistent with \eq{e1e2}. Their values mostly fall between $10^{-3}$ and $10^{-2}$. It suggests that if we observe a sub-Neptune planet pair with relatively high eccentricities ($e\sim 0.1$), they are not likely to be in first-order resonance irrespective of their near-resonance period ratio. Post-disc perturbations \citep{ChoksiChiang2022} could, however, excite eccentricities and change resonant pairs from apsidal anti-alignment to alignment \citep{LauneEtal2022}. These apsidally aligned systems would have slightly larger eccentricities than what \fg{ecc_dist} predicts. 


\begin{figure*}
    \centering
    \includegraphics[width=\columnwidth]{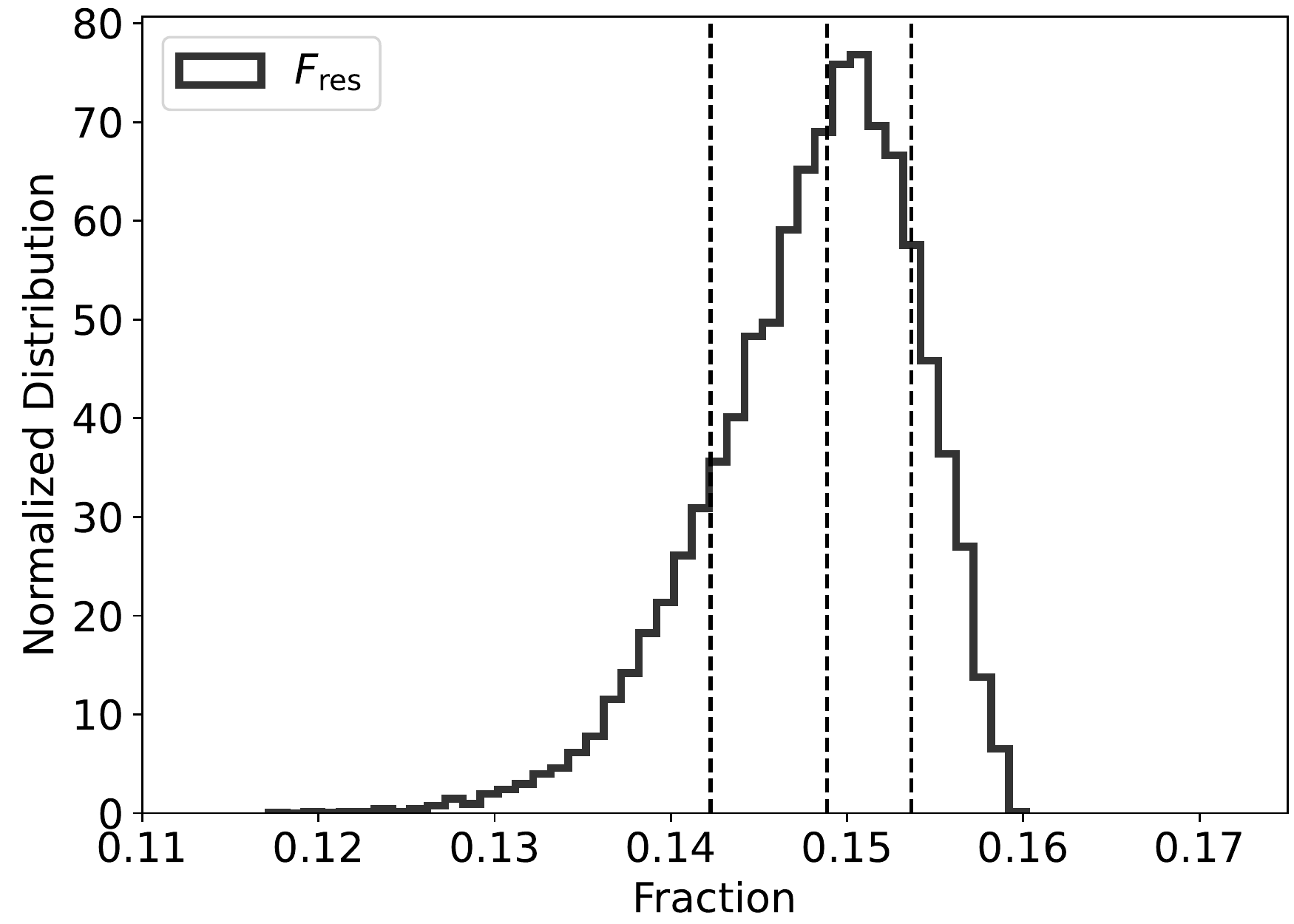}
    \includegraphics[width=\columnwidth]{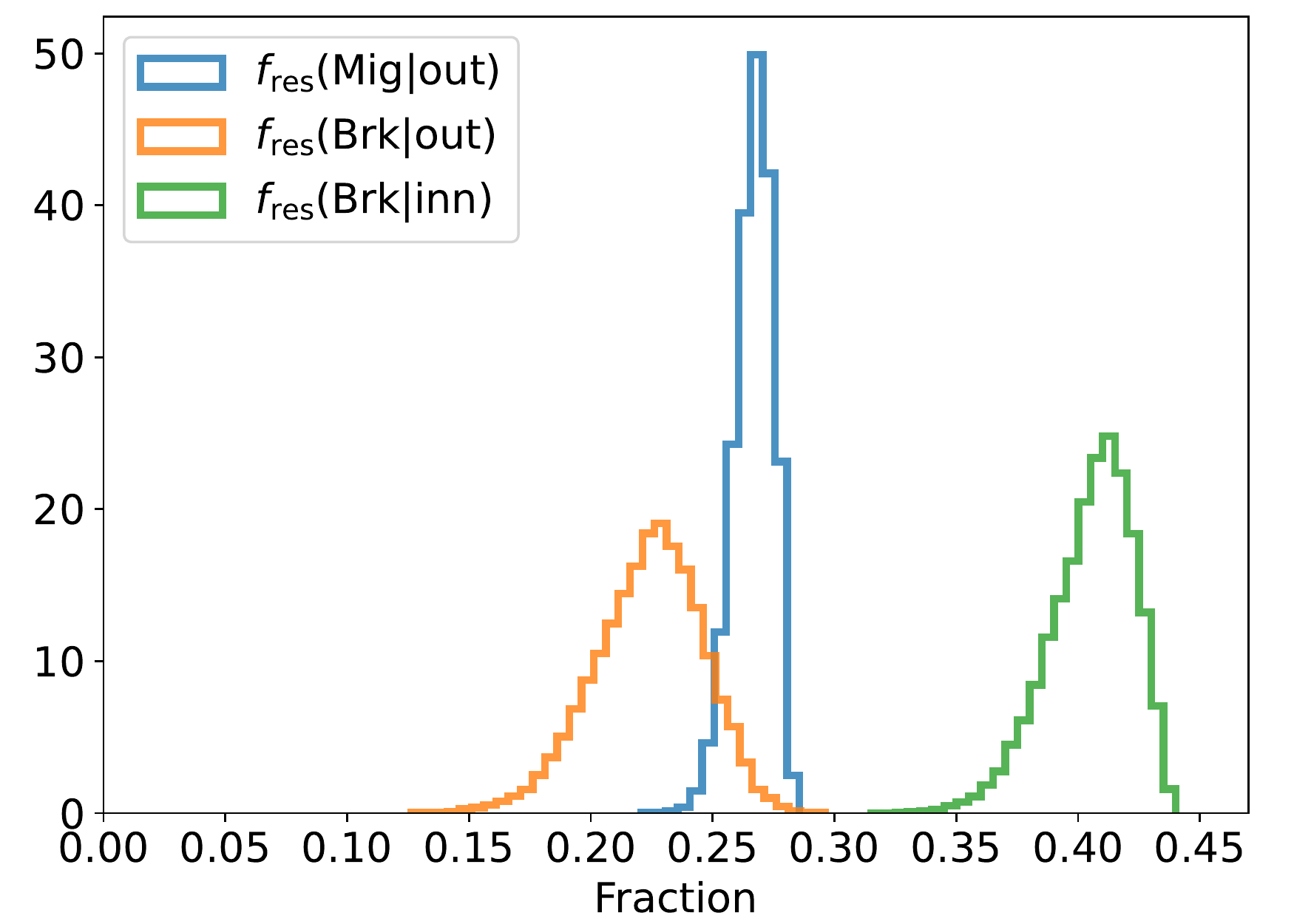}
     \caption{Resulting distribution of resonant fraction for all pairs in our sample (left) and detailed resonant distribution for pairs with period ratio within the near integer ratio windows (right). Here $f_\mathrm{res}(\mathrm{Mig|out)}$ is the fraction of planets that enter resonance during migration with the outer planet migrating faster. $f_\mathrm{res}(\mathrm{Brk|out)}$ is the fraction of resonant planets stopping at the disc inner edge with the outer planet migrating faster. $f_\mathrm{res}(\mathrm{Brk|inn)}$ is the fraction of resonant planets stopping at the disc inner edge with the inner planet migrating faster. }
    \label{fig:res_frac}
\end{figure*}

\subsection{Upper limit on the disc surface density at resonance trapping}
\label{sec:upper_limit}
The constraints the MCMC model provides cannot be used to determine the absolute value of the natal disc surface density, as it cancels in the $\tau_a/\tau_e$ expression. An upper limit for the natal disc surface density of resonant planets can, however, be deduced from the resonance trapping criterion.  
We use \eq{new_criterion} in order to break the degeneracy and to find an upper limit for the disc surface density at the trapping location. 
It is assumed in the derivation of \eq{new_criterion} that the outer planet is on a fixed circular orbit. Such an assumption is valid because the outer planets on average have lower eccentricities than their inner siblings (\fg{ecc_dist}). 
From \se{mcmc_main}, we already found that $\tau_{a}/\tau_{e}>10^3$ for observed transiting planets, that is, overstability is not likely to occur. In this regime, \eq{new_criterion} alone gives the resonance trapping condition. 
Therefore, we can get the critical migration time-scale, below which the resonance would be crossed. In the Type~I migration regime, planet migration speed is proportional to gas disc surface density. We are therefore able to obtain the upper limit of the disc gas surface density.

The upper limit is reached by combining \eq{migration}, \eq{te}, \eq{te1_ta1_s} and \eq{new_criterion}:
\begin{equation}
\label{eq:upper_limit_general}
    \begin{aligned}
        \Sigma_\mathrm{max}(r_2) = & 1.6|f_1|h^3(r_2)\left[(j+1)C_e\gamma_I\frac{\tau_{e_1}}{\tau_{e_2}}\right]^\frac{1}{2}\left(\frac{r_2M_1}{r_1M_2}\right)^\frac{5}{4} 
        \frac{M_\star}{r_2^2}
    \end{aligned}
\end{equation}
inserting a disc model and extrapolating to 1 au, we obtain:
\begin{equation}
\label{eq:upper_limit}
    \begin{aligned}
        \Sigma_\mathrm{1au,max} = & 1.6|f_1|h_\mathrm{1au}^3\left[(j+1)C_e\gamma_I\right]^{{1}/{2}}\left(\frac{M_1}{M_2}\right)^{{3}/{4}} \\
        & \times \left(\frac{r_1}{r_2}\right)^{2q-{s}/{2}-1.5}\left(\frac{r_2}{1\,au}\right)^{3q-s}\frac{M_\star}{r_2^2}.
    \end{aligned}
\end{equation}
The expression $\Sigma_\mathrm{1au, max}\propto h_\mathrm{1au}^3$ is consistent with what \cite{KajtaziEtal2022} found in their simulations. 
In \fg{loc_barrier}, we plot the $\Sigma_\mathrm{1au, max}$ for each planet pair that is possible in resonance, versus their host mass. The deeper the colour, the more likely it is that the pair is in resonance. The size of the symbol indicates the resonance index $j$. The figure shows that high-$j$ resonances tend to be associated with high surface densities. This result is in line with Type-I migration theory. Massive discs result in faster migration, which allows the planets to cross the relatively strong resonance. 
In addition, \fg{loc_barrier} shows that $\Sigma_\mathrm{max}$ increases with stellar mass. That is because the migration speeds in the Type I limit depend on the star-to-disc mass ratio. Higher surface density is required to migrate faster. 
We indicate the surface density of the Minimum mass solar nebula \citep[i.e., MMSN,][]{Hayashi1981} and the Minimum mass extra-solar nebula \citep[i.e., MMEN,][]{ChiangLaughlin2013} in \fg{loc_barrier}. The upper limits of nearly all disc surface densities are below that of the MMEN, while the inferred disc surface densities are centred around the MMSN value. 

We conclude this discussion with two final points. First, the value for $\Sigma_\mathrm{max}$ we obtained refers to the time when the planets were locked into resonance, not the upper limit over the entire disc lifetime. Second, planets in higher $j$ resonances tend to provide higher upper limits on the disc surface density. However, they might alternatively have formed in close proximity to each other, avoiding crossing of lower-$j$ resonances. In that case, the true value of $\Sigma_\mathrm{1au}$ is likely to be less than $\Sigma_\mathrm{1au,max}$. 

\subsection{Migration barrier reflects the disc inner rim}
\label{sec:barrier}
Planets can migrate in the disc, but their migration is believed to be halted somewhere as otherwise all planets would be consumed by the host star. However, the location and mechanism of the migration barrier are under debate. There are two main explanations for the barriers: dust sublimation \citep{KamaEtal2009, FlockEtal2019} and the stellar magnetosphere \citep{ KoeniglEtal2011, HartmannEtal2016}. 

If we assume a mass-luminosity relation for main sequence stars \citep{Duric2004}, the dust sublimation radius becomes:
\begin{equation}
    r_\mathrm{subl} = r_\mathrm{sub,0}\left(\frac{M_\star}{M_\odot}\right)^2,
\end{equation}
where $r_\mathrm{sub,0}$ is the silicate dust sublimation radius for solar mass stars:
\begin{equation}
\label{eq:sublimation}
    r_\mathrm{sub,0}=0.13\left(\frac{C_\mathrm{bw}}{1}\right)^{1/2} \left(\frac{0.1}{\epsilon}\right)^{1/2} \left(\frac{1400 \mathrm{K}}{T_\mathrm{d}}\right)^{2} \mathrm{au},
\end{equation} 
where $C_\mathrm{bw}$ is the back-warming factor, $\epsilon$ is dust cooling efficiency and $T_\mathrm{d}$ is the dust sublimation temperature \citep{KamaEtal2009}. 

\begin{figure}
    \centering
    \includegraphics[width=\columnwidth]{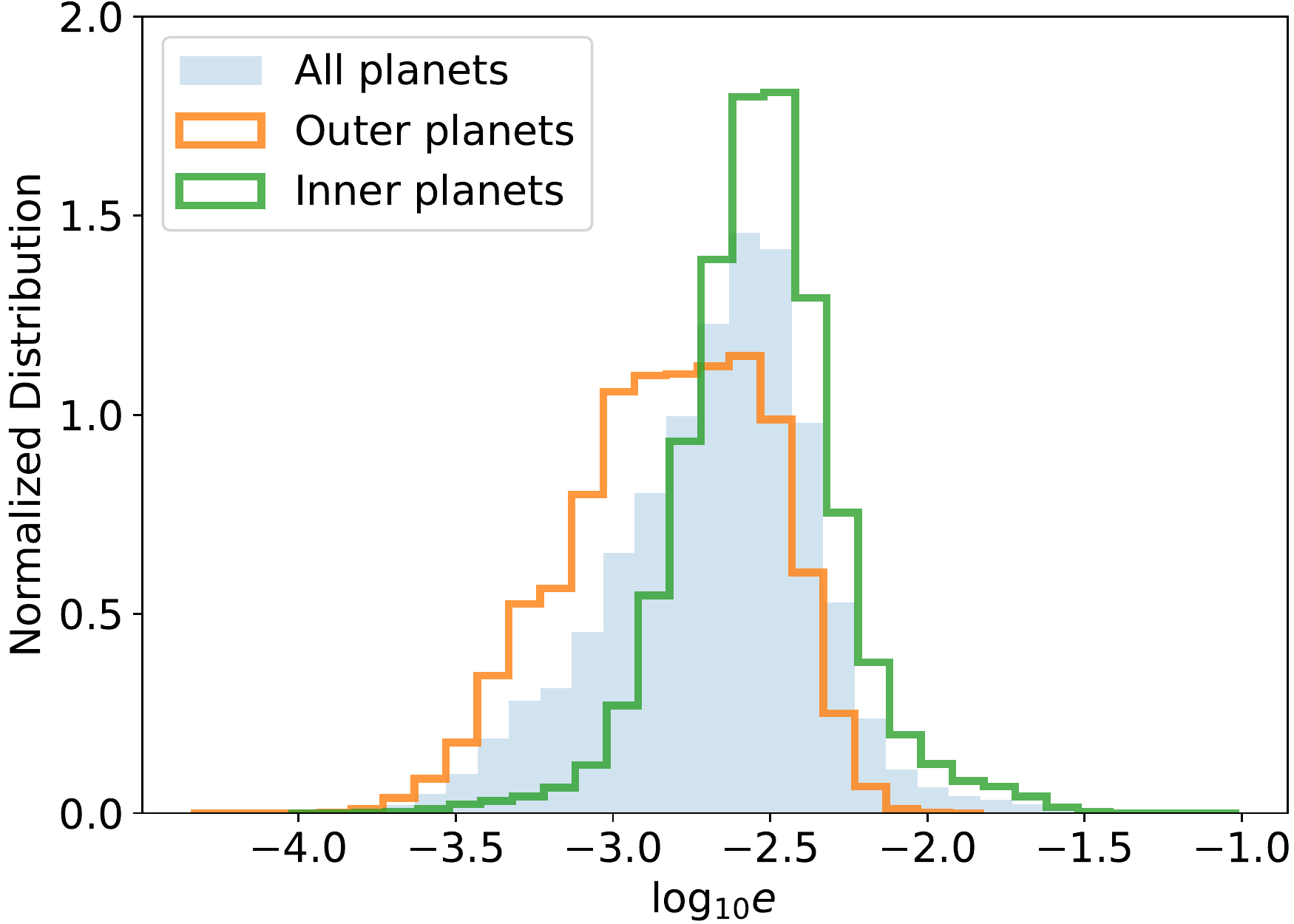}
     \caption{Eccentricity distribution of planet pairs that are in resonance, weighted by the probability of being in resonance. We compare the eccentricity distribution of the outer planets in all planet pairs with their inner planets. }
    \label{fig:ecc_dist}
\end{figure}

\begin{figure*}
    \centering
    \includegraphics[width=1.8\columnwidth]{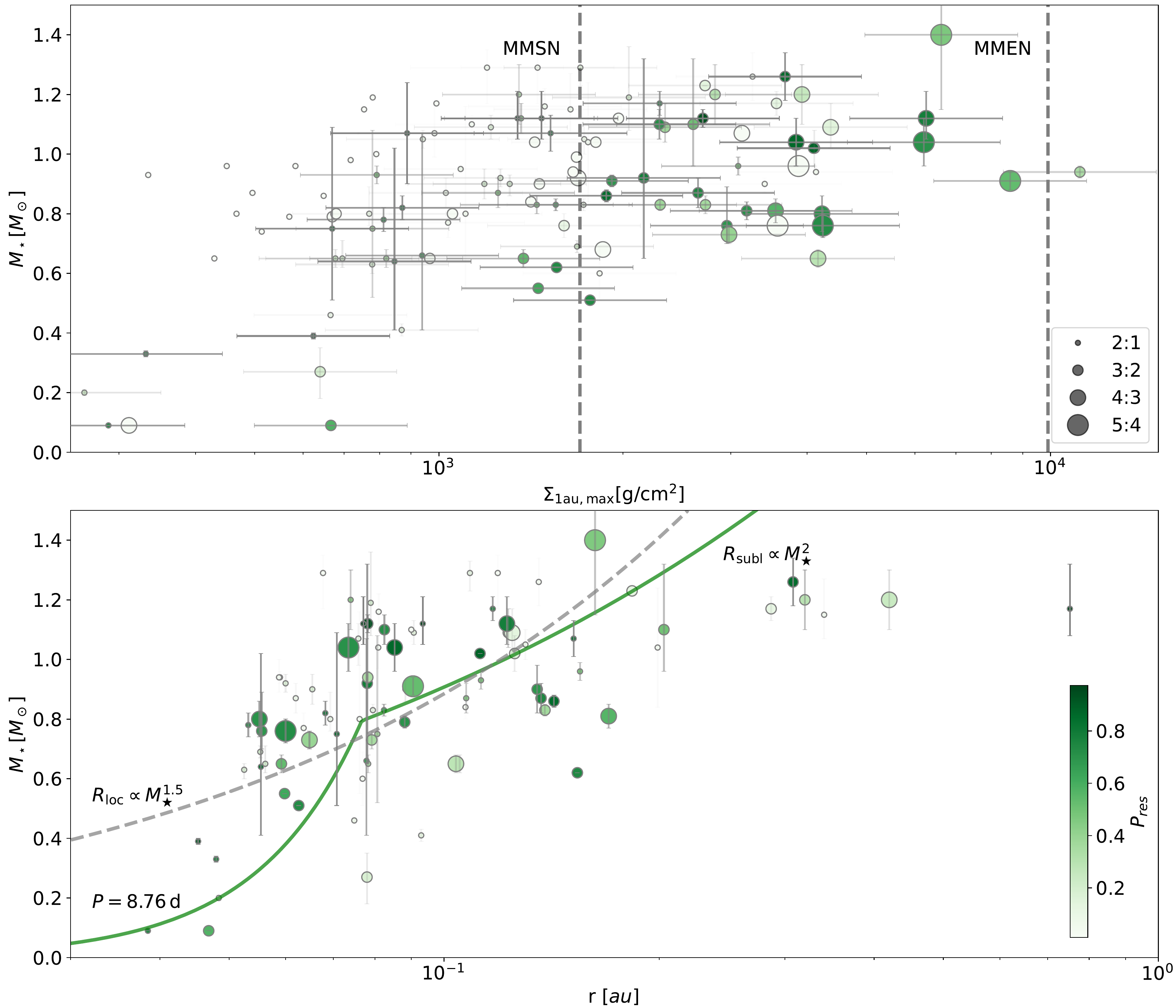}
    \caption{Inferred maximum surface density at 1~au for the resonant planets (top panel) and location of the migration barrier (bottom panel) versus the host stellar mass{, for the 128 planet pairs in the four windows in \fg{per_dist}}. Different sizes of the circles indicate different resonance index $j$. The colour of the symbols indicates the probability of planets in resonance ($P_{\mathrm{res},k}$). In the top panel, error bars indicate the 1$\sigma$ uncertainties in the stellar mass and $\Sigma_\mathrm{1au, max}$ -- the latter follows from the uncertainty in $C_e$. The left and right grey dashed lines indicate the surface density of the MMSN and MMEN, respectively. In the bottom panel, grey and green lines give the best-fitting location of the migration barrier (the location of the inner planet in resonant pairs) fitted by a single power law and a broken power law, respectively. }
    \label{fig:loc_barrier}
\end{figure*}

The magnetospheric infall radius tends to expand and converge to the stellar corotation radius due to angular momentum locking \citep{LongEtal2005}. The stellar corotation radius is expressed as:
\begin{equation}
    r_\mathrm{co}=0.057\left(\frac{P_\star}{5\, \mathrm{d}}\right)^{\frac{2}{3}}\left(\frac{M_\star}{M_\odot}\right)^{\frac{1}{3}} \mathrm{au}
\end{equation}
The rotation periods of T Tauri stars \citep[e.g.][]{BouvierEtal2007, LeeChiang2017} and young star associations \citep[e.g.,Upper Sco and NGC 2264][]{RoquetteEtal2021} are several days (1-10 days). 

The MCMC model is capable to identify (in a probabilistic sense) whether or not the planets have reached a migration barrier.
We plot the location of the inner planet of those pairs -- the presumed location of the migration barrier -- in \fg{loc_barrier}. The symbol size again represents the resonance index $j$ and the colour indicates how likely they are in resonance. From the plot, it can be seen that as the stellar mass increases the location of the migration barrier moves further away from the star. Motivated by the two theories about the migration barrier and aiming to figure out which radius is more consistent with observation, we fit the location of a migration barrier using the single power law expression: 
\begin{equation}
\label{eq:barrier}
    r_{\mathrm{bar},k} = r_0 \left(\frac{M_{\star,k}}{M_\odot}\right)^l, 
\end{equation}
where $r_0$ is the location of the migration barrier for solar-mass stars. The Gaussian likelihood we construct is weighted by the resonance probability $P_{\mathrm{res},k}$:
\begin{equation}
    \begin{aligned}
        \ln\mathcal{L} = \sum_{k=1}^{N_\mathrm{pairs}}\ln\left[ \mathcal{N}( r_{\mathrm{bar},k}-r_k, \sigma_r^2)\cdot P_{\mathrm{res},k}\right],
    \end{aligned}
\end{equation}
where $r_{\mathrm{bar},k}$, $r_k$ and $\sigma_r$ are the model predicted migration barrier location for the $k$-th pair (\eq{barrier}), and the presumed location of the migration barrier and the standard deviation of the fitted migration barrier radius, respectively. If one pair has a larger $P_{\mathrm{res},k}$, it is more likely that the planets are formed in the protoplanet disc and undergo disc migration. The MLE fits $r_0=0.13$, $l=1.49$ and $\sigma_r$ is 0.079 au. The location of the migration barrier fitted by single power law relation is, however, shallower than the corresponding index of the dust sublimation radius but steeper than that of the magnetospheric radius. It may imply that there are two planet populations whose migration barriers are carved by either dust sublimation radius or magnetospheric radius. For this reason, we also fit a broken power law:
\begin{equation}
    r_{\mathrm{bar},k}=
    \displaystyle
    \left\{\begin{array}{ll}
        r_1 \left(\frac{M_{\star,k}}{M_\odot}\right)^{1/3} & \text{if } M_{\star,k} \le M_\mathrm{crit}\\
        r_2 \left(\frac{M_{\star,k}}{M_\odot}\right)^{2} & \text{if } M_{\star,k} > M_\mathrm{crit},
    \end{array}
    \right.
    \label{eq:broken_law}
\end{equation}
where $M_\mathrm{crit}=\left(r_1/r_2\right)^{3/5}M_\odot$ is the transition mass between the two different power laws. Similar to the single power law, we use MLE to fit the two parameters $r_1=0.08\,\mathrm{au}$ and $r_2=0.12\,\mathrm{au}$ and the transition mass $M_\mathrm{crit}$ is $0.79M_\odot$ (see \fg{loc_barrier}). The low mass fit corresponds to a corotation period of $P_\star=8.76$ days, which is consistent with the observed rotation period for young stars \citep{BouvierEtal2007, LeeChiang2017, RoquetteEtal2021}. The value of $r_2$ agrees with \eq{sublimation}. Around these high-mass stars, planets are trapped at the dust sublimation location as it exceeds $r_\mathrm{co}$. 
However, the uncertainty $\sigma_r = 0.078$ au is only slightly smaller than the value given by the single power law, suggesting that the broken power law model is only marginally better. A larger sample will make for a more reliable analysis. 

\section{Discussion}
\label{sec:discussion}
In this work, we have constructed a model that connects the planet migration history to the observed values of the offset from integer period ratios ($\Delta$). If the resulting planet pair is in resonance, $\Delta$ follows a log-normal distribution. On the other hand, if it is non-resonant, the corresponding $\Delta$ is assumed to follow a uniform distribution. Based on this, we have developed a statistical model that constrains the migration histories of the observed planets by conducting a Markov Chain Monte Carlo method (MCMC) analysis. We examine our MCMC method using self-generated mock data \se{model_test} and prove that it can indeed reproduce certain features.

Our model for resonance trapping is designed for two-planet systems and first-order resonances, but we also include observed systems with planets in a resonance chain. \cite{KajtaziEtal2022} have shown that the averaged properties of the resonance chain (involving three or more planets) still reflect the properties of the system as if there is only one resonant pair. Therefore, multi-planet systems do not significantly contaminate the results.
In addition, we have not considered higher-order resonances. A more general model that applies to both first-order and higher-order resonances needs to be considered in the future. 
Finally, it is possible that the inner planets migrate across the inner disc edge and enter the disc cavity \citep{HuangOrmel2022i, FitzmauriceEtal2022i}, where our model would not be applicable. But those planets are plausibly massive enough to open a deep gap \citep{AtaieeKley2021, ChrenkoEtal2022}, which are excluded by our sample selection.

A key assumption in the model is that the uncertainties in the planet masses and the ensuing $\Delta$ follow a log-normal distribution. We can then fit the excess of period ratio just wide of integer ratio with a log-normal profile, thus extracting pairs in resonance. If we would adopt different distributions for the mass, the resulting distributions for $\Delta$ would become far more complex and no longer allow us to express the corresponding likelihoods in closed form. 
However, certain post-disc dynamics, e.g., post-disc energy dissipation from planetesimal scattering \citep{ChatterjeeFord2015, GhoshChatterjee2022}, stellar tides \citep{LithwickWu2012, BatyginMorbidelli2013} and stellar encounters \citep{CaiEtal2019, StockEtal2020}, could slightly change the period ratios of planet pairs. 
Therefore, they may play roles in broadening, shifting, or even skewing the log-normal profile. What the resulting $\Delta$ distribution may look like needs to be investigated in future work. Once addressed, one may learn the post-disc perturbation histories the planets have experienced. However, this also requires a much larger sample size than what we have at present, as already in this work the MCMC is unable to break some model degeneracies. 



Our model mainly applies to those small planets unable to open a gap (in Type~I migration regime). \cite{TanakaEtal2002} and \cite{TanakaWard2004} showed that the semi-major axis damping time-scale ($\tau_{a}$) and eccentricity damping time-scale ($\tau_{e}$) have a relation for locally isothermal disc: $\tau_{e}=1.28C_eh^2\tau_{a}/\gamma_I$ and $C_e=1$. If the planet partially opens a gap, the migration speed decreases linearly with the gas surface density in the gap \citep{KanagawaEtal2018}. The same holds for eccentricity damping \citep{PichierriEtal2022}. Therefore, the ratio $\tau_{a}/\tau_{e}$ is independent of surface density and the constraints we obtain on it still hold when the planet opens a partial gap. 
Neither assuming a specific disc nor migration model, we obtain that $\tau_a/\tau_e\approx10^4$, which is the most robust result of this study. After adopting the irradiation disc model, we further obtain $C_e\approx1$ which is consistent with \cite{TanakaEtal2002} and \cite{TanakaWard2004}. On the other hand, \cite{CharalambousEtal2022} argue that $C_e=0.1$ because lower $C_e$ increases the resonance offsets. A potential reason is that their simulations have assumed that trapping takes place at 1 au, where the disc aspect ratio is relatively large, while most transiting planets are found at ${\sim}0.1$ au. 
Furthermore, by improving the analytical criterion for resonance trapping, we are able to constrain the upper limit on the natal disc surface density for those planets in resonance. 
The resulting maximum surface density is similar to that of the Minimum Mass Solar Nebular (MMSN) but smaller than that of the Minimum Mass Extra-solar Nebula (MMEN). 

{Several other migration prescriptions have been proposed, including more sophisticated ones such as those by \cite{PaardekooperEtal2010, PaardekooperEtal2011}. These prescriptions demonstrate that planets within a certain mass range can be naturally trapped at a location where the (positive) corotation torque exerted on the planet exceeds the (negative) Lindblad torque \citep{BitschEtal2013, BitschEtal2014, BaruteauEtal2014}. To first order, for small planets, the migration behavior would still be predominantly linear (with planet mass and disk mass) except near these trapping locations. In our model, such a scenario is naturally incorporated through the \stalling pair. Regarding the dependence of the damping terms on planet eccentricity, non-linear correction terms have been proposed by e.g., \cite{CresswellNelson2006, CresswellNelson2008, IdaEtal2020}. However, at low eccentricity, these non-linear terms are irrelevant and we do not include these terms in our investigation. Finally, our findings on the ratio $\tau_a/\tau_e$ are robust, irrespective of the specific disc migration prescriptions and non-linear terms. This is because this ratio has a one-to-one correspondence to the resonance offset $\Delta$. Therefore, our conclusion regarding $\tau_a/\tau_e$ remains valid even if we incorporate different migration models (disc structures) or non-linear terms.}

The imprint resonant trapping leaves behind tentatively allow us to assess where and when planets form in the discs. 
The evaluation of the probability of planet pairs in MMR is crucial, and this step also informs us of the number of resonant pairs. Since we employ different models for how planets get trapped in resonance (\se{MCMC_model}), it is possible to distinguish gas-poor formation scenarios (pairs trapped in resonance when migrating) and gas-rich formation scenarios (pairs stopped by the migration barrier). The former could be identified with late formation, while the latter, which are more dominant, are connected to the early formation in gas-rich discs. The present orbits of these planets further hint at the location of migration barriers. However, due to our small sample, we cannot unambiguously identify the physical origin of the migration barrier; either the dust sublimation radius or the magnetospheric radius would fit the data. The situation is, however, expected to improve in the near future. The upcoming launch of the PLAnetary Transits and Oscillations of stars (PLATO) mission \citep[e.g.][]{RauerEtal2014} and The Earth 2.0 (ET) mission \citep{GeEtal2022, Ye2022} could drastically increase not only the number but also the precision of planet detections. It will provide us with a more precise analysis of the planet formation and migration histories reflected in the dynamical properties of resonant planets.

\section{Conclusions}
\label{sec:conclusion}
We manage to construct a statistical model connecting planet migration theory to observed quantities of Kepler planets. Based on the inferred masses and resonance offsets, we conduct an MCMC analysis to extract the history of planet-disc interaction from planet-planet dynamics. The statistical approach provides us with the following findings:
\begin{enumerate}
    \item The {semi-major axis-to-eccentricity damping time-scale ratio} can be constrained at $\log_{10} \tau_a/\tau_e \approx 4$ with a dispersion of ${\approx}0.3$ dex, irrespective of the assumed disc model. The eccentricity damping is so efficient that overstable libration of resonances is unlikely to have occurred. 
    \item After assuming the isothermal irradiation disc model, we obtain that  $\log_{10}C_e=0.24^{+0.25}_{-0.25}$, which connects eccentricity damping and semi-major axis damping in the Type-I migration theory. It is consistent with \cite{TanakaEtal2002} and \cite{TanakaWard2004}. 
    \item From the MCMC posterior, the probability that a planet pair is in resonance follows. The fraction of transit planet pairs in first-order MMR amounts to $14.8_{-0.7}^{+0.5}$\%.
    \item Most of the inferred resonant planets are consistent with the scenario that they reached a migration barrier, indicative of early migration in a gas-rich disc. The location of the migration barrier could be the dust sublimation radius for massive stars ($M_\star>0.79M_\odot$) and the magnetospheric radius for low mass stars ($M_\star<0.79M_\odot$). 
    \item By evaluating the resonance strength of those inferred resonant planets, the upper limit of the proto-disc surface density during the planet formation era is obtained. Most systems have their $\Sigma_\mathrm{1au}$ below that of the Minimum Mass Extra-solar Nebula (MMEN) and half of them below that of the Minimum Mass Solar Nebula (MMSN). 
    \item It is found that the classical MMR trapping/crossing criterion based on the pendulum model does not match numerical simulation. We provide and numerically verify an improved criterion (\eq{new_criterion}) based on the equilibrium resonance angle, which together with the overstability condition of \citet{GoldreichSchlichting2014} (\eq{overstable}) fully describes the problem.
\end{enumerate}

Future work could feature a more complete model accounting for high-order resonances and resonance chains. Detection of more resonant planets and more precise measurements of planetary and stellar properties with upcoming missions will definitely improve the confidence of our analysis.


\section*{Acknowledgements}
The authors appreciate the thoughtful comments of the referee, Dr.~Carolina Charalambous. S.H. would like to thank Wei Zhu, Mario Flock, Simon Portegies-Zwart, Martijn Wilhelm, and Remo Burn for their useful discussions. The authors acknowledge support by the National Natural Science Foundation of China (grant no. 12250610189).
This work has made use of the NASA Exoplanet Archive. 
Software: \texttt{emcee} \citep{Foreman-MackeyEtal2013}, \texttt{corner.py} \citep{corner}, \texttt{Matplotlib} \citep{Hunter2007,CaswellEtal2021}, \texttt{REBOUND} \citep{ReinLiu2012} and \texttt{REBOUNDx} \citep{TamayoEtal2020}.

\section*{Data Availability}
The data underlying this article will be shared on reasonable requests to the corresponding author.




\bibliographystyle{mnras}
\bibliography{ads} 




\appendix
\section{Results of the General disc model}
\begin{figure*}
    \centering
    \includegraphics[width=1.5\columnwidth]{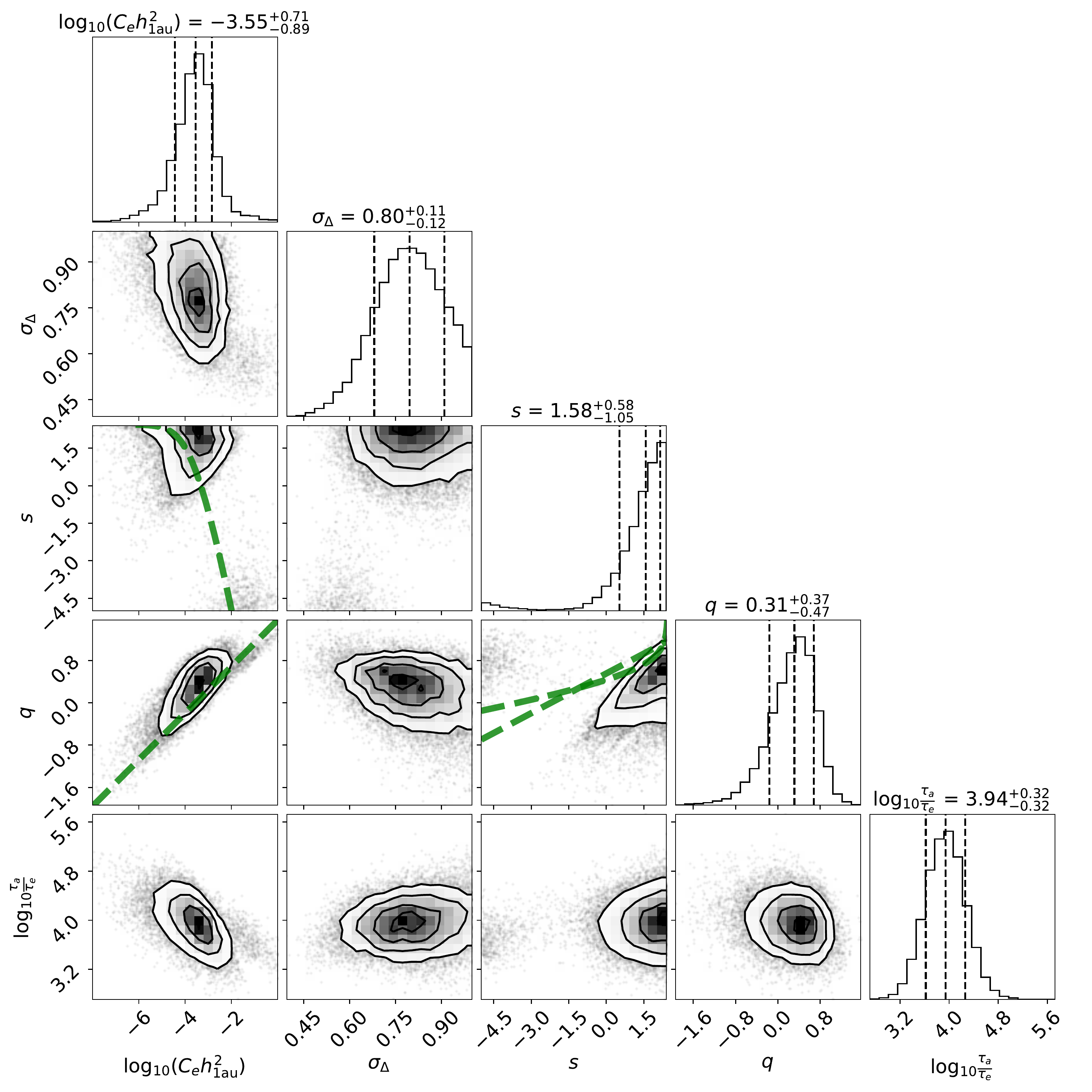}
    \caption{
    Corner plot of variables in the MCMC analysis ($\log_{10}C_e$, $\sigma_\Delta$, $s$, and $q$) with 1$\sigma$, 2$\sigma$ and 3$\sigma$ confidence contours, w/o assuming the disc structure. We also calculate $\log_{10}(\tau_a/\tau_e)$ accordingly and add this variable into the corner plot. {The 1$\sigma$ uncertainty is labelled on the top of each column and indicated by left and right dashed lines. The middle dashed lines indicate their median values.} Green lines indicate the correlation between different variables, which is \eq{equi1} and \eq{equi2}. 
    }
    \label{fig:mcmc_corner_general}
\end{figure*}

\label{sec:mcmc_general}
The full MCMC fitting corner plot of the General model is shown in \fg{mcmc_corner_general}. The fitted values for all four parameters are $\log_{10}(C_eh_\mathrm{1au}^2)=-3.55^{+0.71}_{-0.89}$, $s=1.58^{+0.58}_{-1.05}$, $q=0.31^{+0.37}_{-0.47}$. However, these values do not necessarily refer to the real physical disc parameters, due to the degeneracy among them. Therefore, we generate mock planet period ratio data (with the same size as our sample used here) and examine this MCMC model without specifying the disc structure in \se{model2_test_general}. Those tests indeed suggest that $s$ and $q$ are not able to be retrieved.

\fg{mcmc_corner_general} reveals the degeneracy: $s$ negatively correlates with $\log_{10}(C_eh_\mathrm{1au}^2)$, $q$ correlates positively with $\log_{10}(C_eh_\mathrm{1au}^2)$ and $q$ positively with $s$. We found that such correlations are well represented by:
\begin{equation}
\label{eq:equi2}
    4q-s=C_\mathrm{2}, 
\end{equation}
from \eq{te1_te2} and 
\begin{equation}
\label{eq:equi1}
    \log_{10}(C_eh_\mathrm{1au}^2)+2q\log_{10}\frac{r_2}{1\,au}-\log_{10}(0.78\gamma_I) +(4q-s)\log\frac{r_1}{r_2}=C_\mathrm{1},
\end{equation}
from \eq{te1_ta1_m}, \eq{te1_ta2} and \eq{te1_ta1_s}. $C_\mathrm{1}$ and $C_\mathrm{2}$ are two constants.  \eq{equi1} and \eq{equi2} are indicated in \fg{mcmc_corner_general} (green dashed lines), and they match the correlation. The fitted log-normal dispersion $\sigma_\Delta$, however, does not have any correlation with other parameters. It makes sense because our model does not depend on $\sigma_\Delta$.
We introduce another parameter:
\begin{equation}
\label{eq:logtate}
    \log_{10}\frac{\tau_a}{\tau_e}=\frac{C_eh(\overline{r}_1)^2}{0.78\gamma_I} 
\end{equation}
and it can be calculated given $\{\log_{10}(C_eh_\mathrm{1au}^2), \sigma_\Delta, s, q\}$. Here, $\overline{r}_1$ is the observed semi-major axis of inner planet average over all planet pairs. This parameter is calculated and put in the corner plot. It shows that $\log_{10}{\tau_a}/{\tau_e}$ is a quantity independent of $s$ and $q$, but slightly depends on $\log_{10}(C_eh_\mathrm{1au}^2)$. 

\section{MCMC performance examination}
\label{sec:model_test}

\begin{figure*}
    \centering
    \includegraphics[width=\columnwidth]{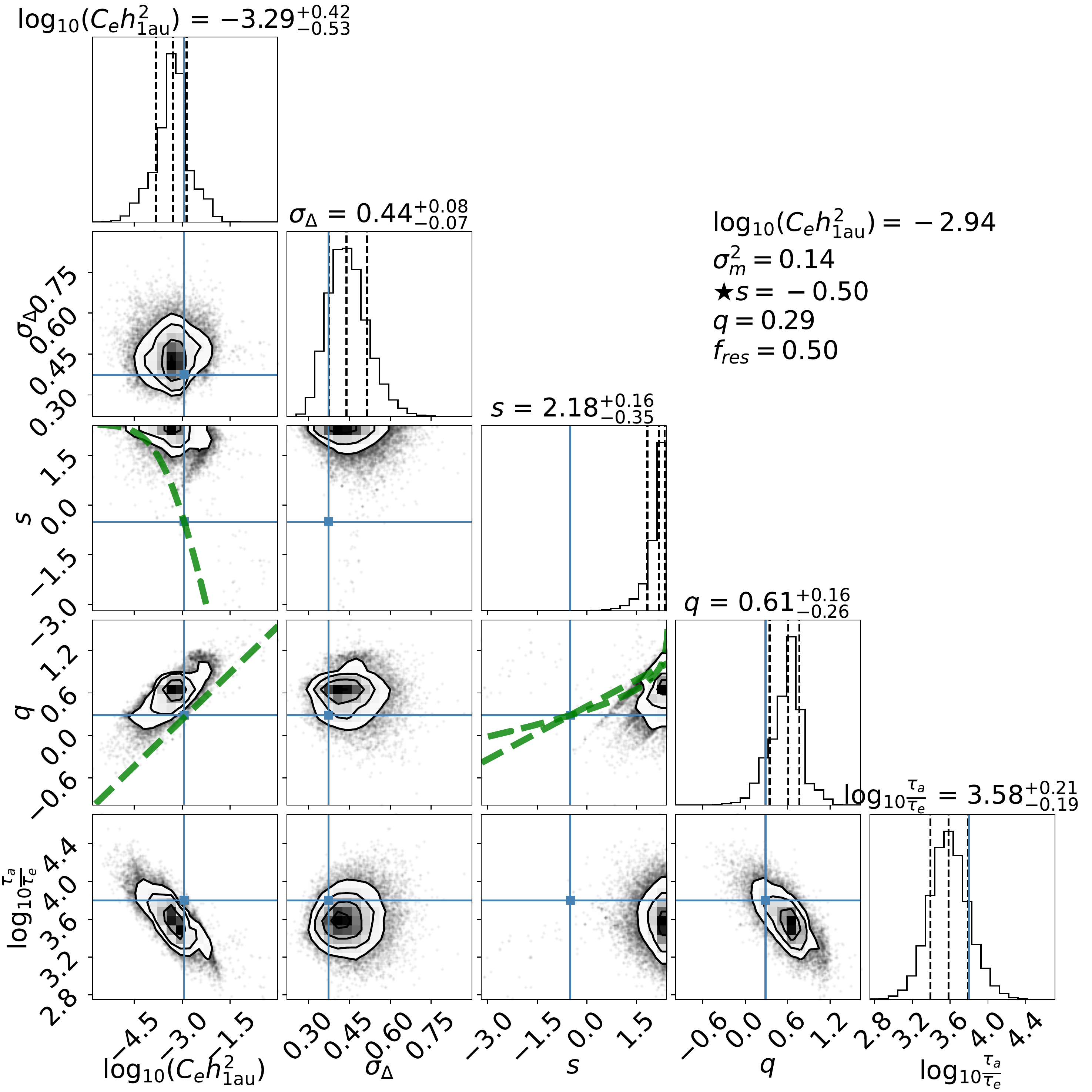}
    \includegraphics[width=\columnwidth]{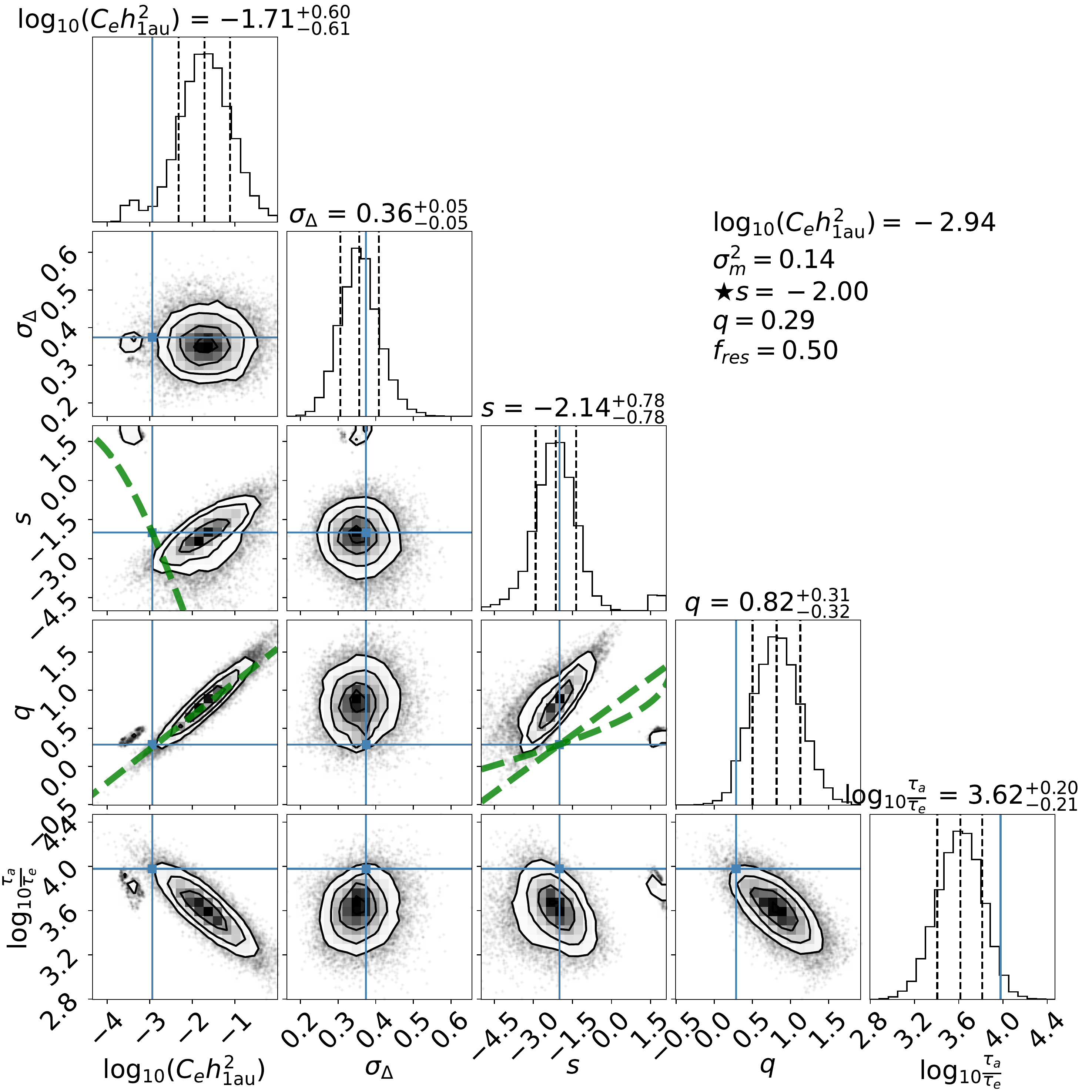}
    
    \caption{Similar to \fg{mcmc_corner_general}. Examination of MCMC model performance, without specifying the disc structure. In each panel, all parameters used are labelled in the upper right corner. We change the parameter $s$ (labelled after $\bigstar$) in different panels to examine the performance of MCMC model and we change $s$ from $-0.5$ and $-2$. In each corner plot, blue lines show the true values. {The 1$\sigma$ uncertainty is labelled on the top of each column and indicated by left and right dashed lines. The middle dashed lines indicate their median values.} }
    \label{fig:model2_separate_s}
\end{figure*}

\begin{figure*}
    \includegraphics[width=\columnwidth]{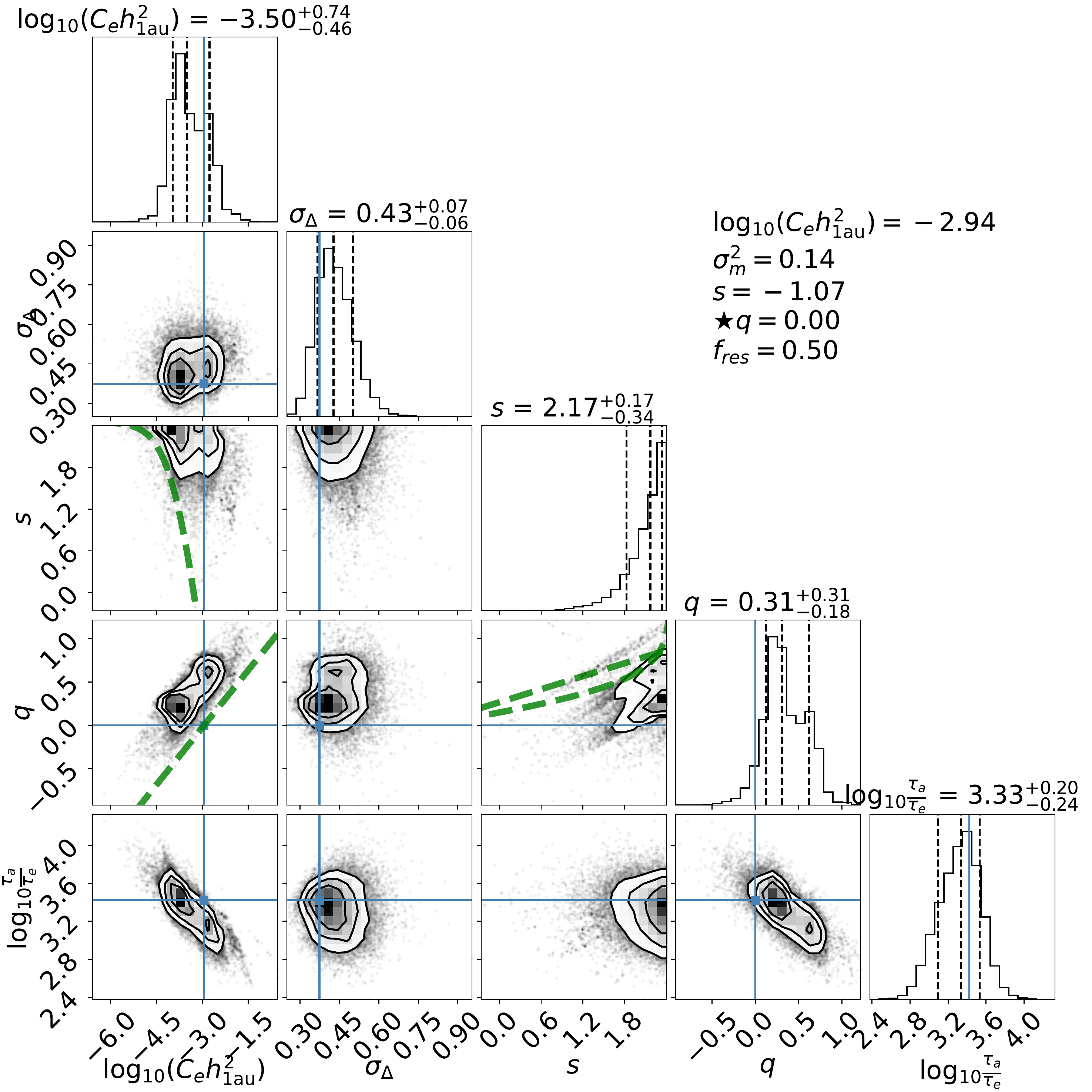}
    \includegraphics[width=\columnwidth]{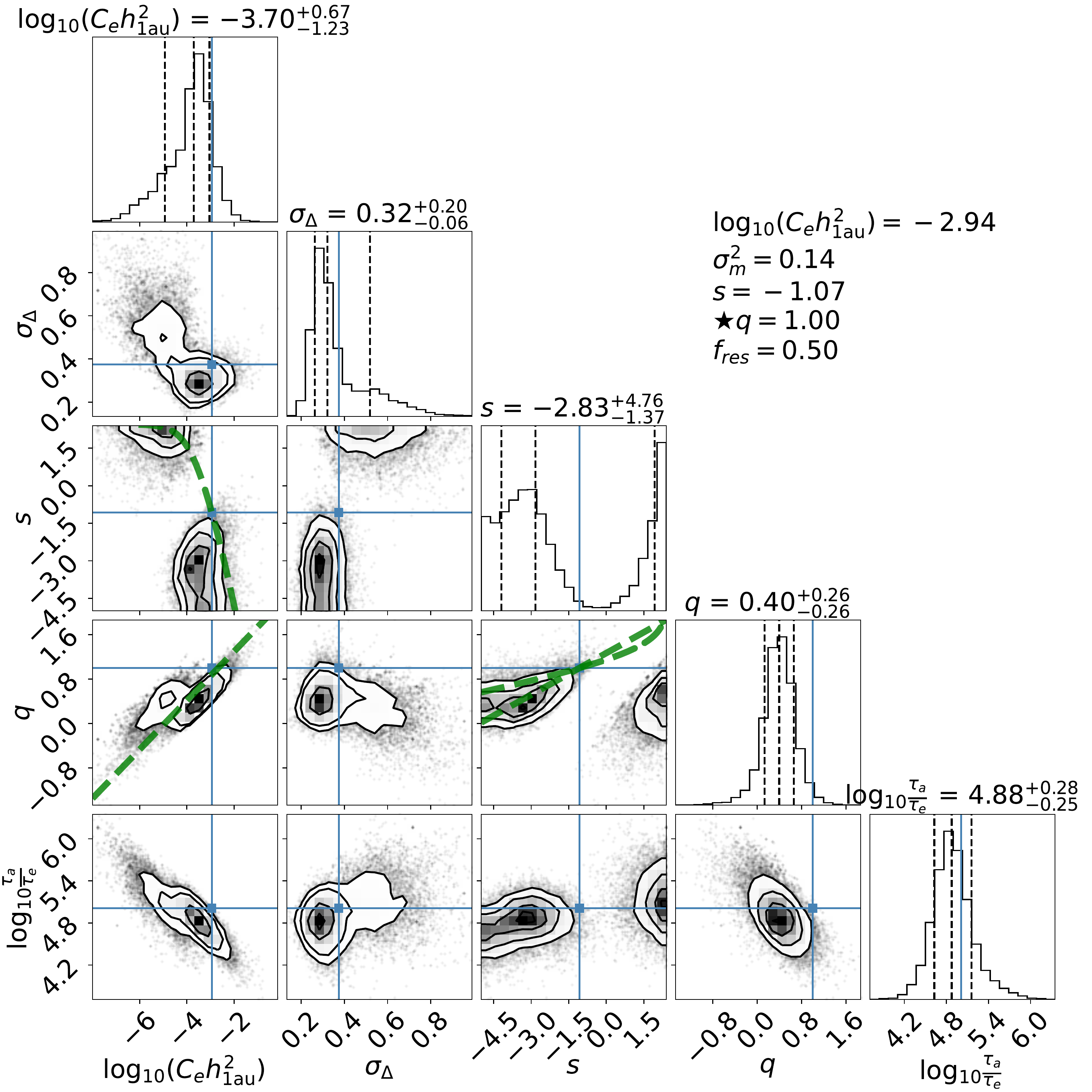}
    \caption{Similar to \fg{model2_separate_s}, but with different $q$. We change $q$ from $0$ and $1$. }
    \label{fig:model2_separate_q}
\end{figure*}

\begin{figure*}
    \includegraphics[width=\columnwidth]{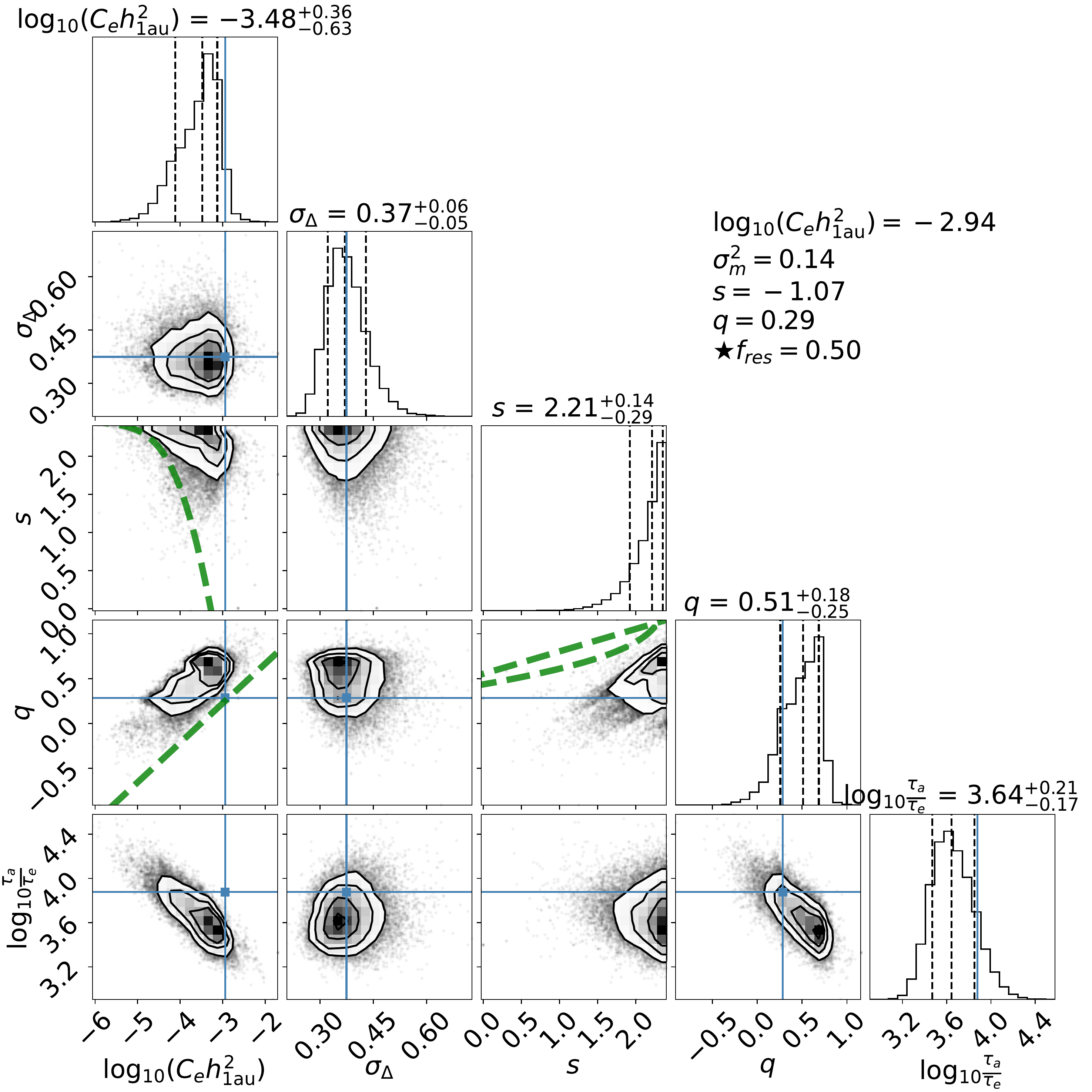}
    \includegraphics[width=\columnwidth]{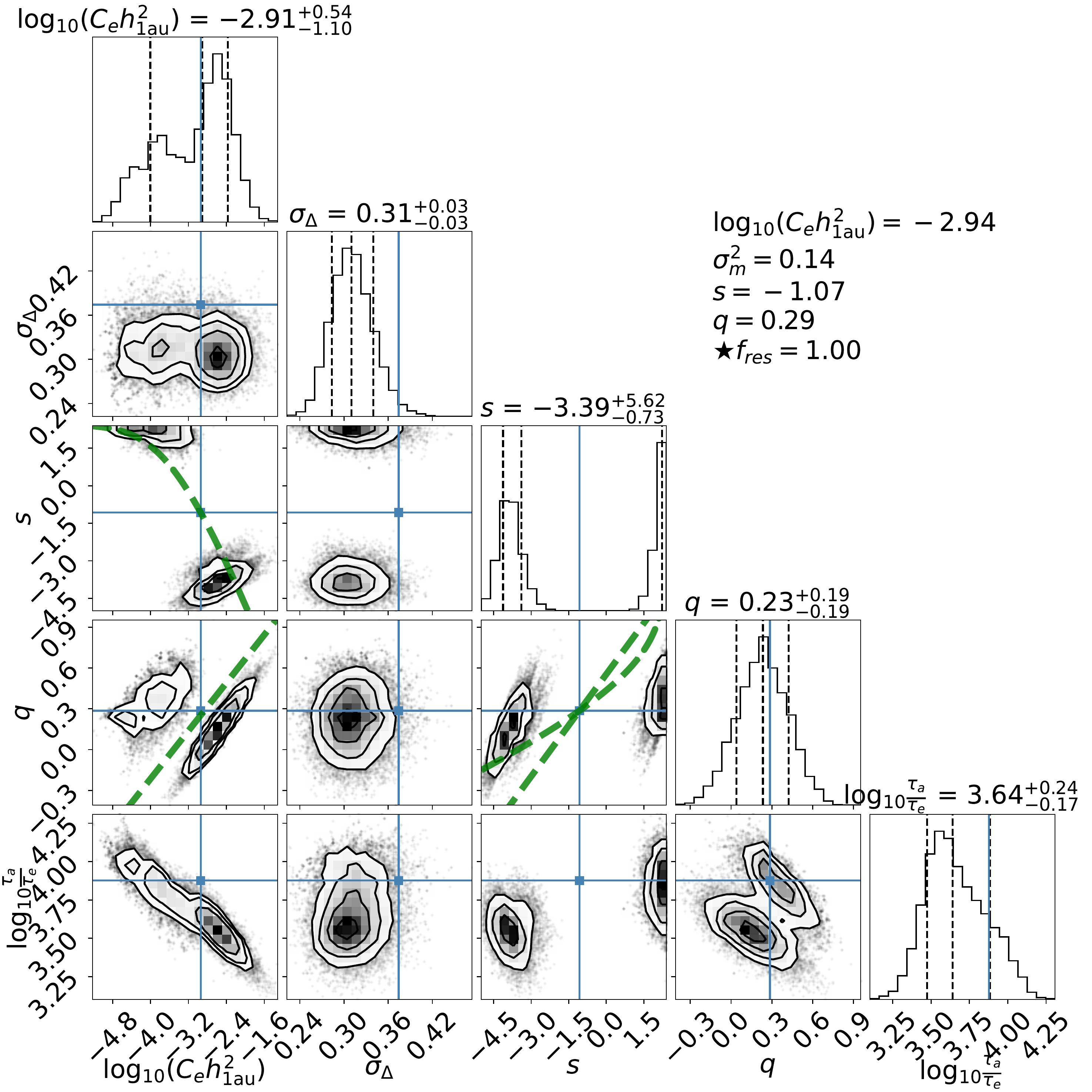}
    \caption{Similar to \fg{model2_separate_s}, but with different $f_\mathrm{res}$. We change fraction of pairs in resonance ($f_\mathrm{res}$) from $50\%$ to $100\%$. }
    \label{fig:model2_separate_fres}
\end{figure*}

\begin{figure*}
    \centering
    \includegraphics[width=0.6\columnwidth]{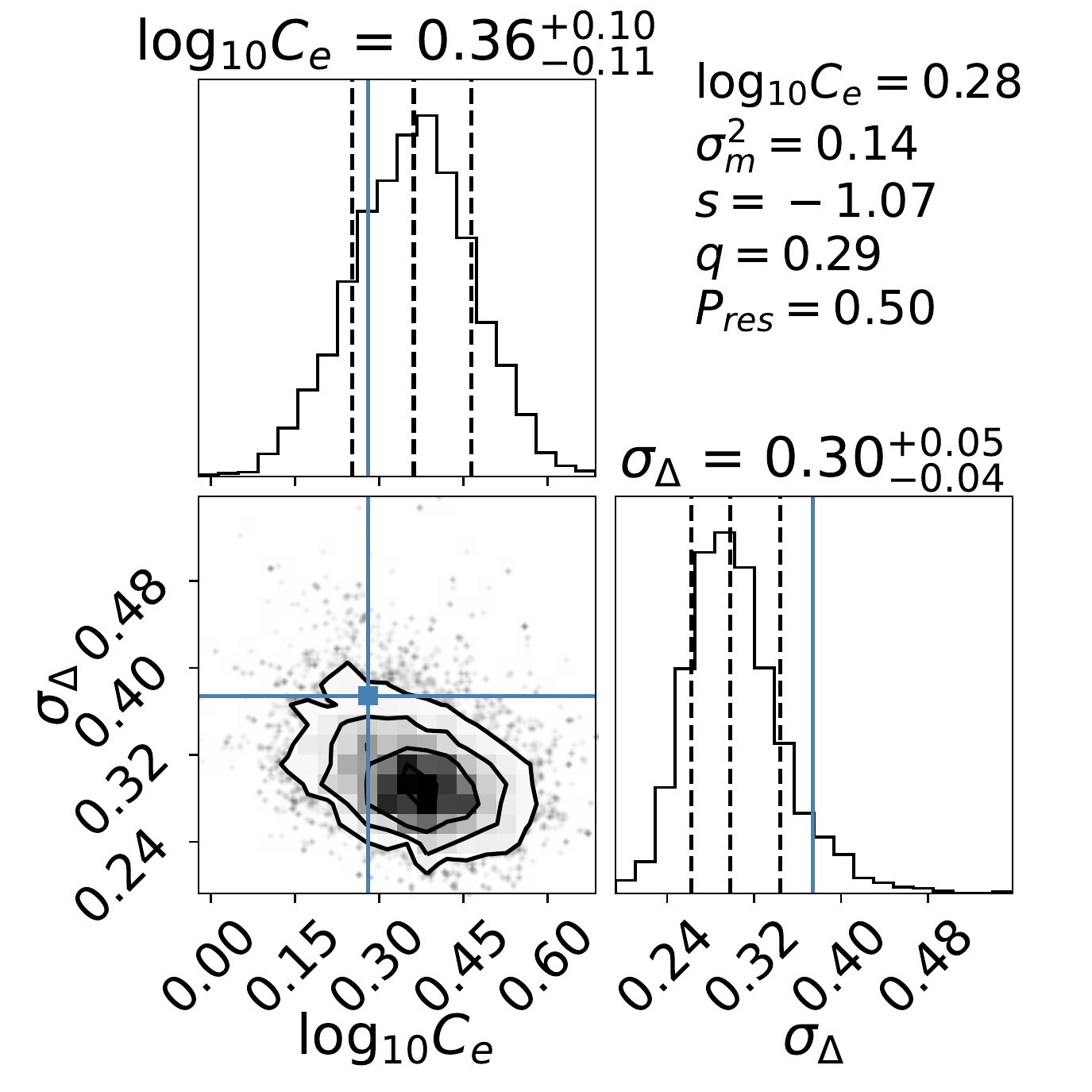}\\
    \includegraphics[width=0.5\columnwidth]{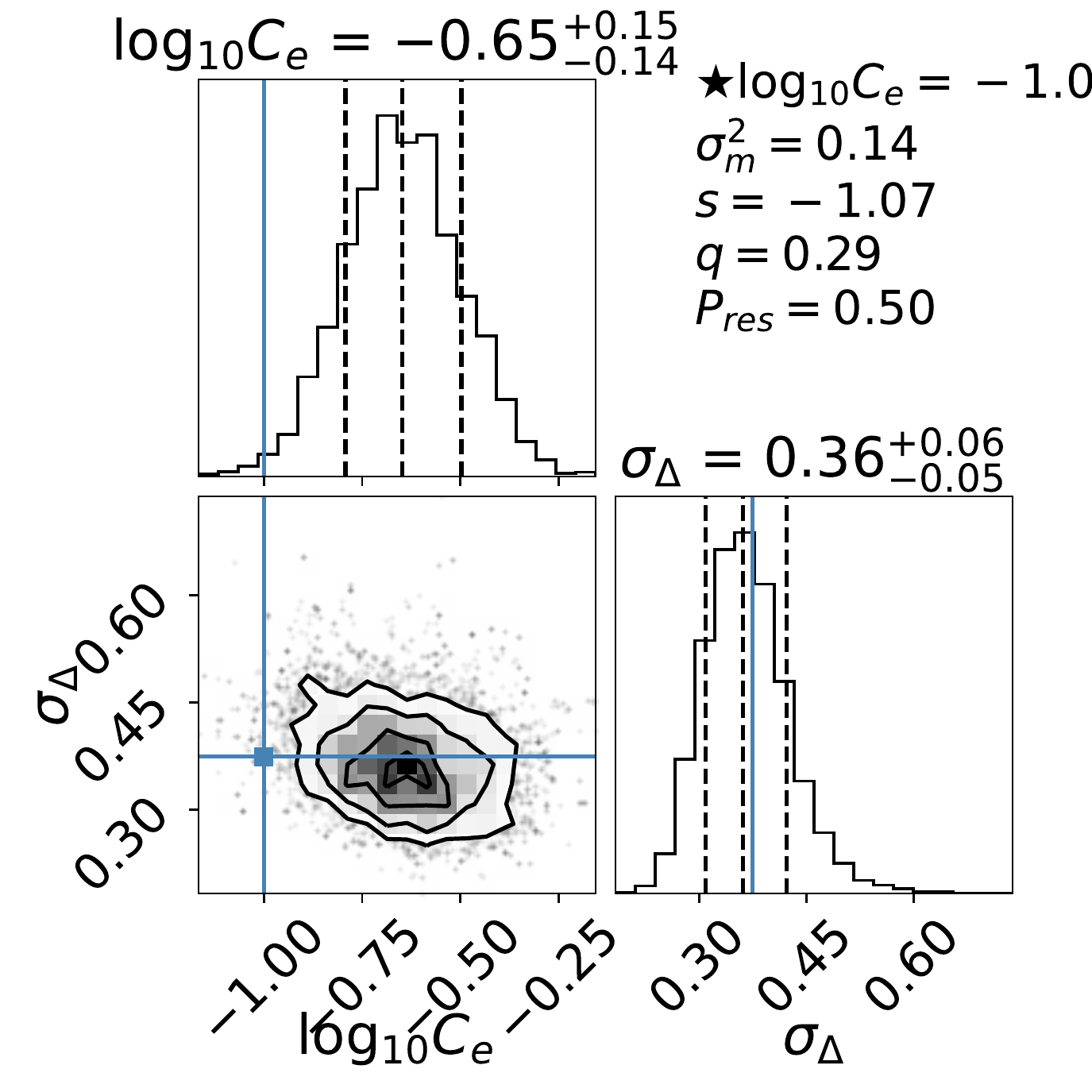}
    \includegraphics[width=0.5\columnwidth]{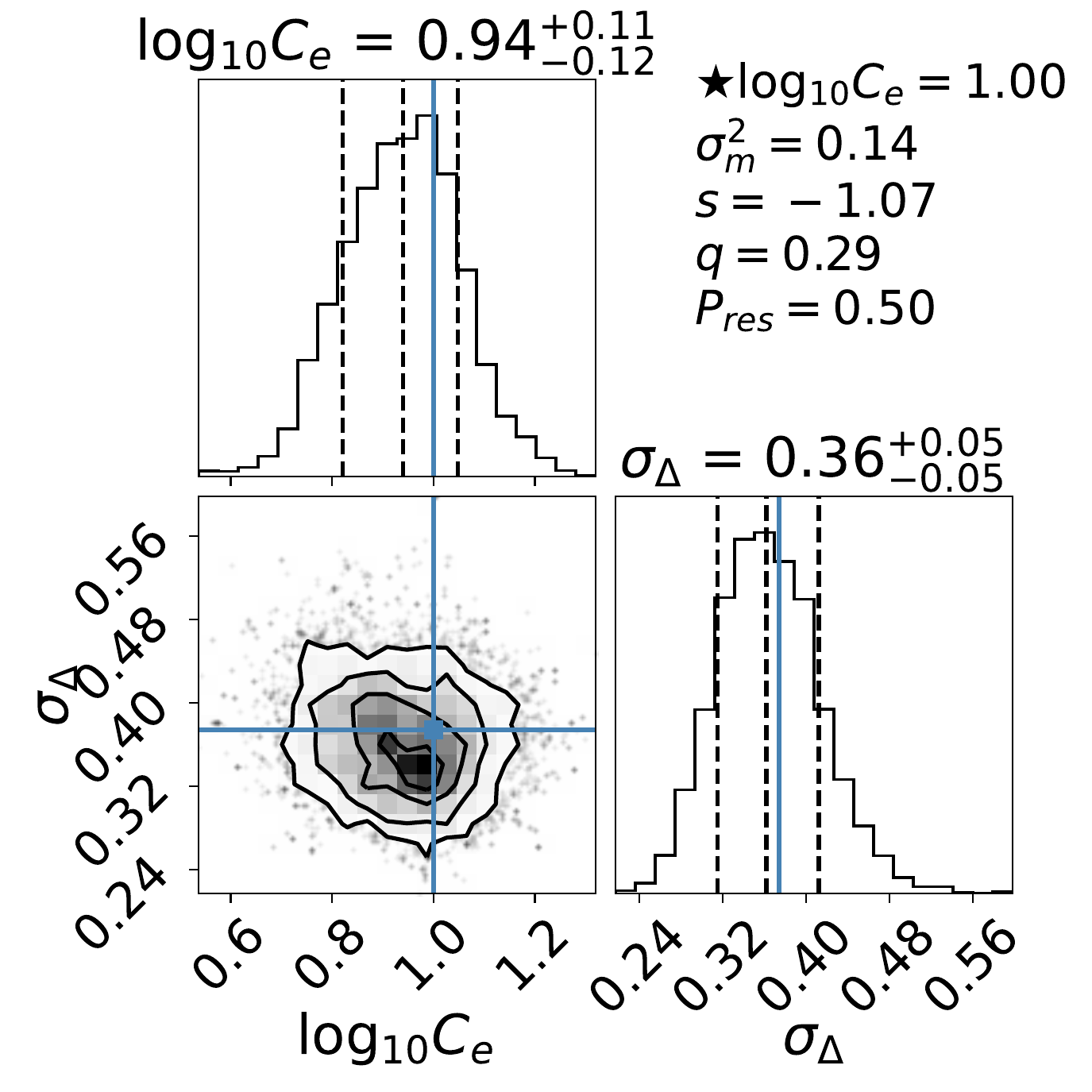}
    \includegraphics[width=0.5\columnwidth]{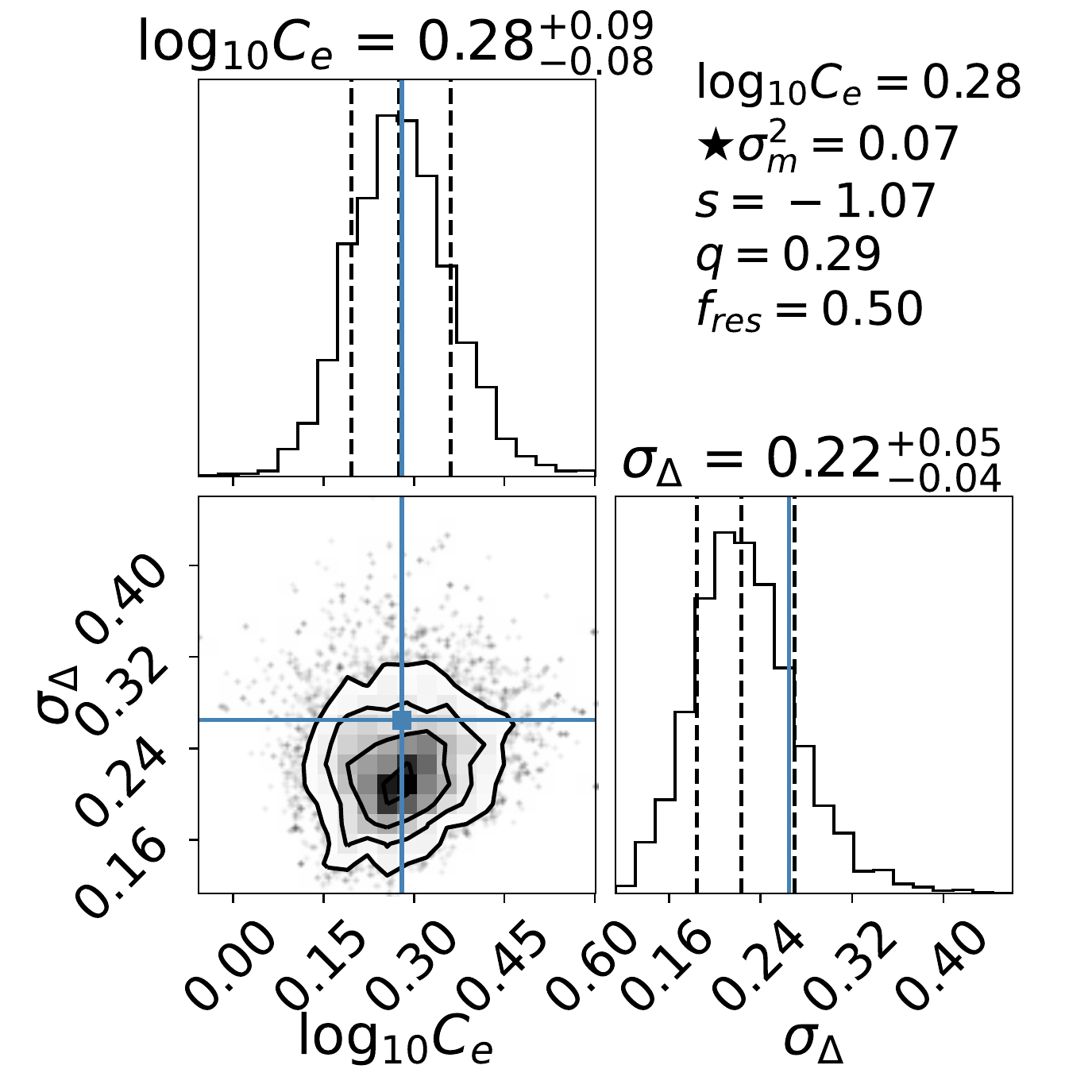}
    \includegraphics[width=0.5\columnwidth]{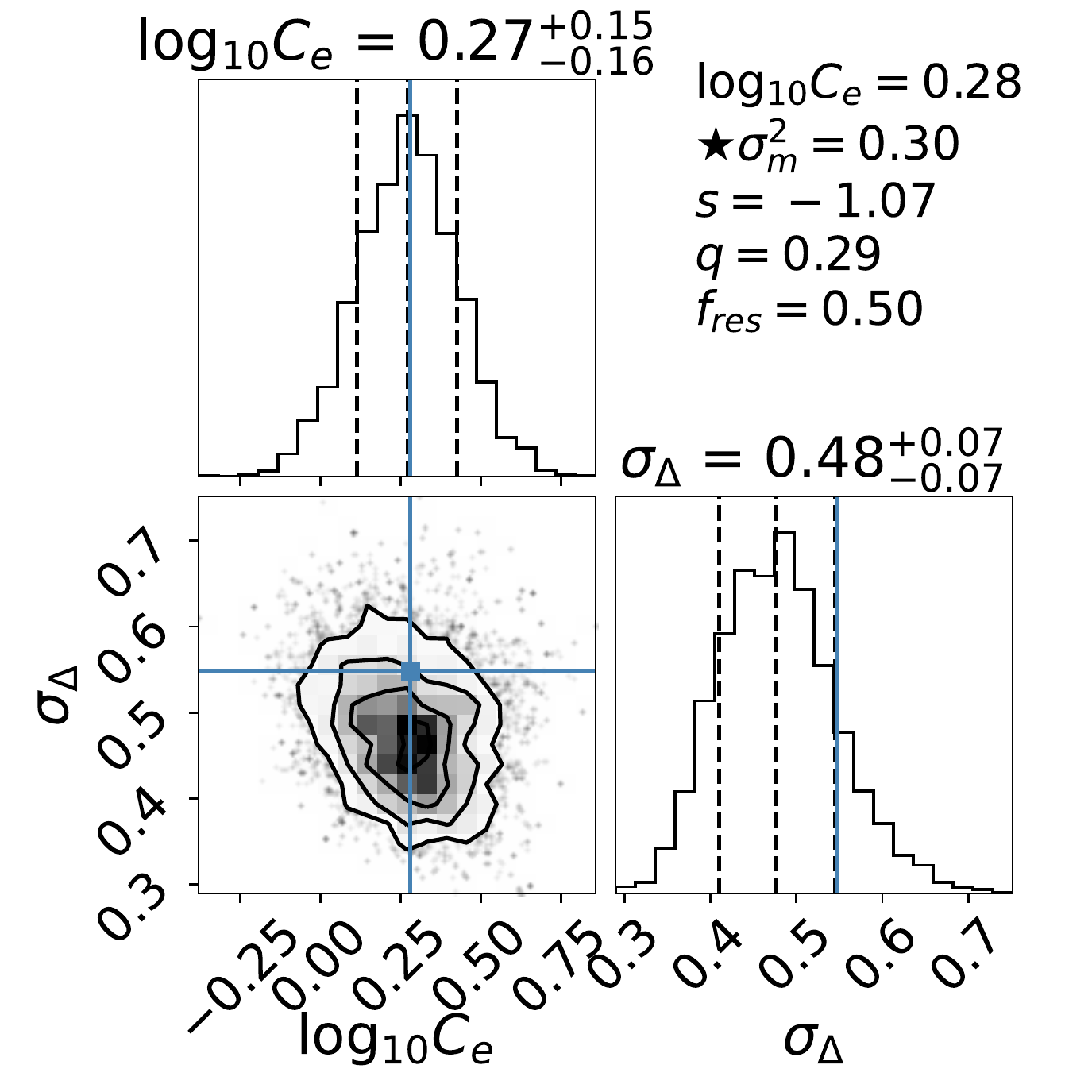}
    \includegraphics[width=0.5\columnwidth]{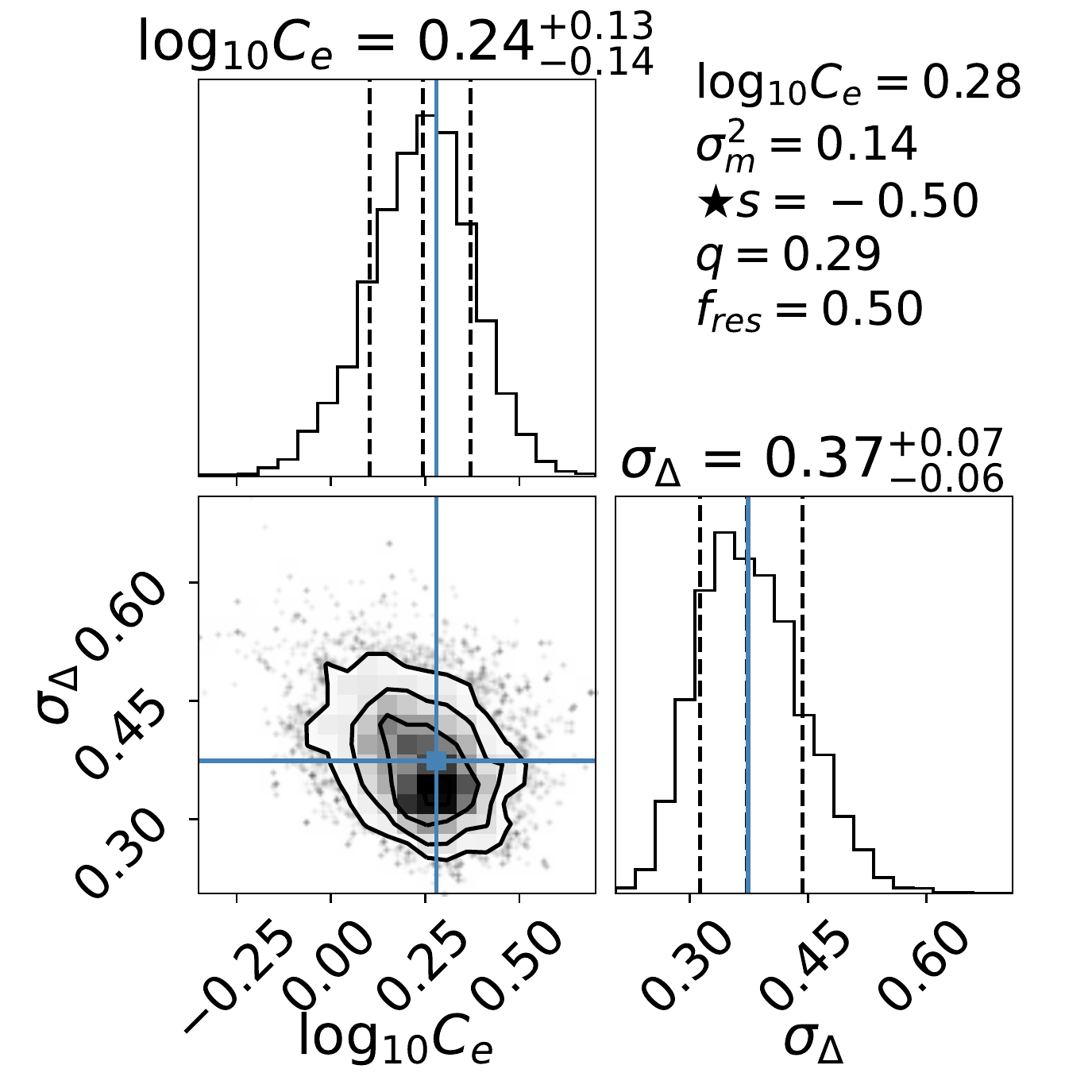}
    \includegraphics[width=0.5\columnwidth]{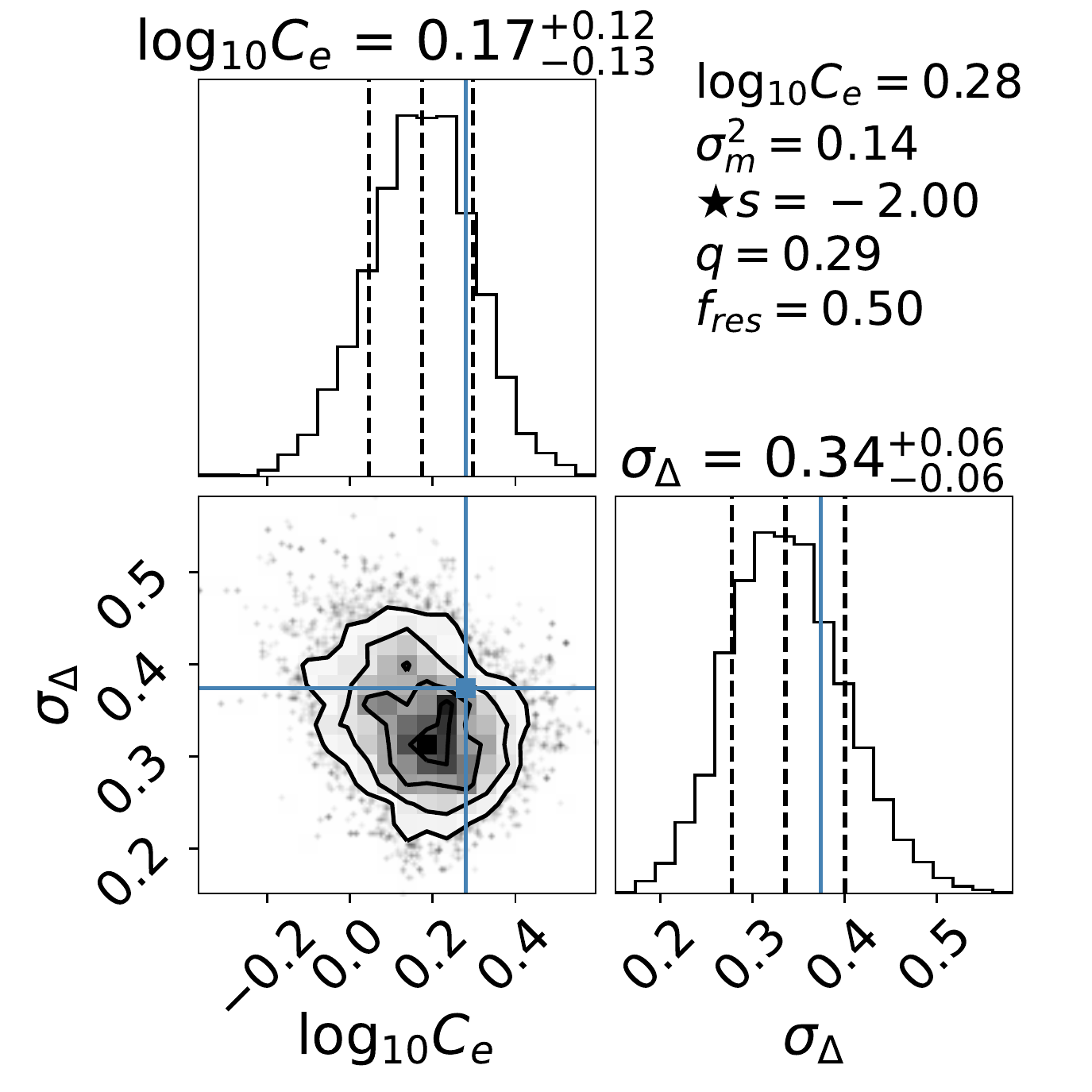}
    \includegraphics[width=0.5\columnwidth]{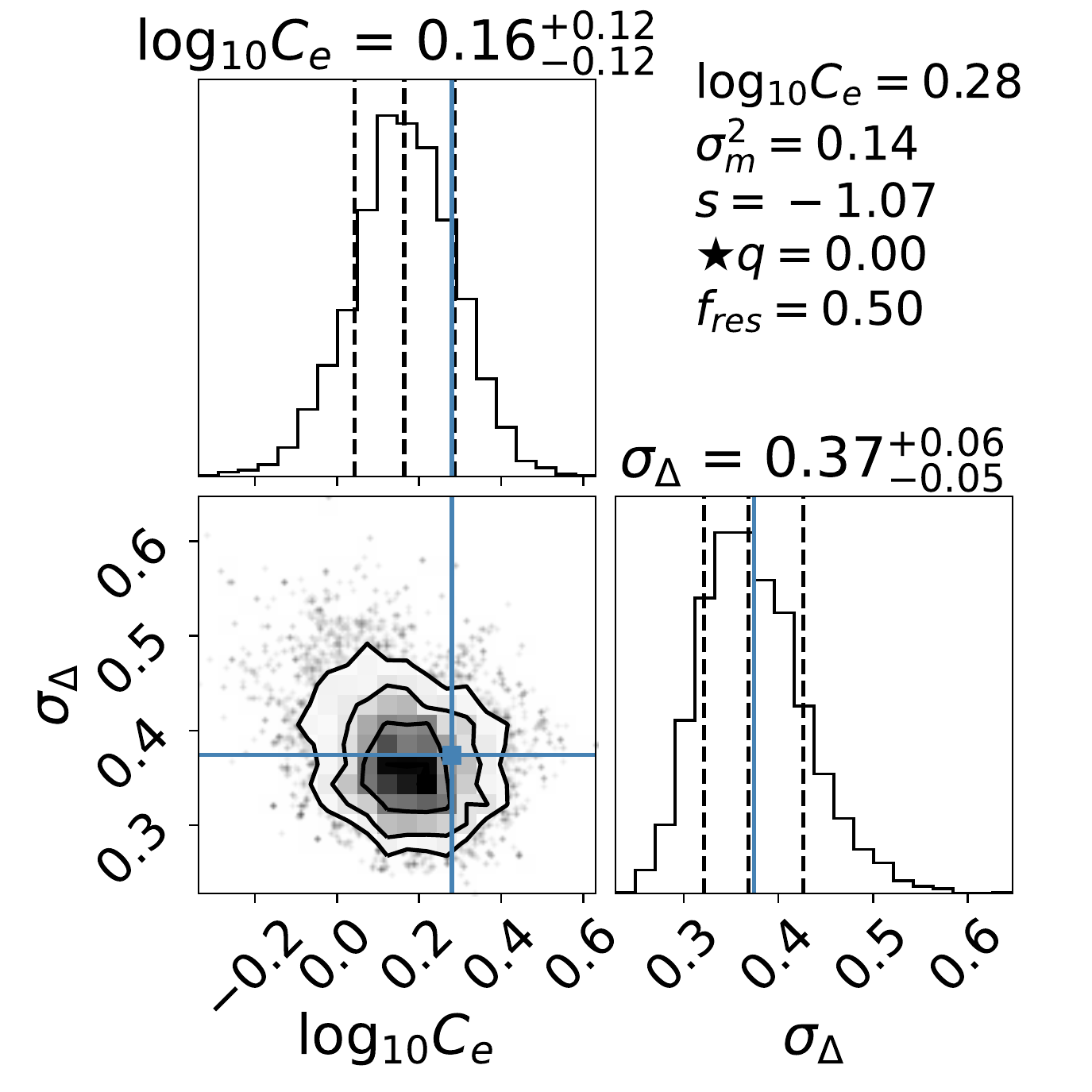}
    \includegraphics[width=0.5\columnwidth]{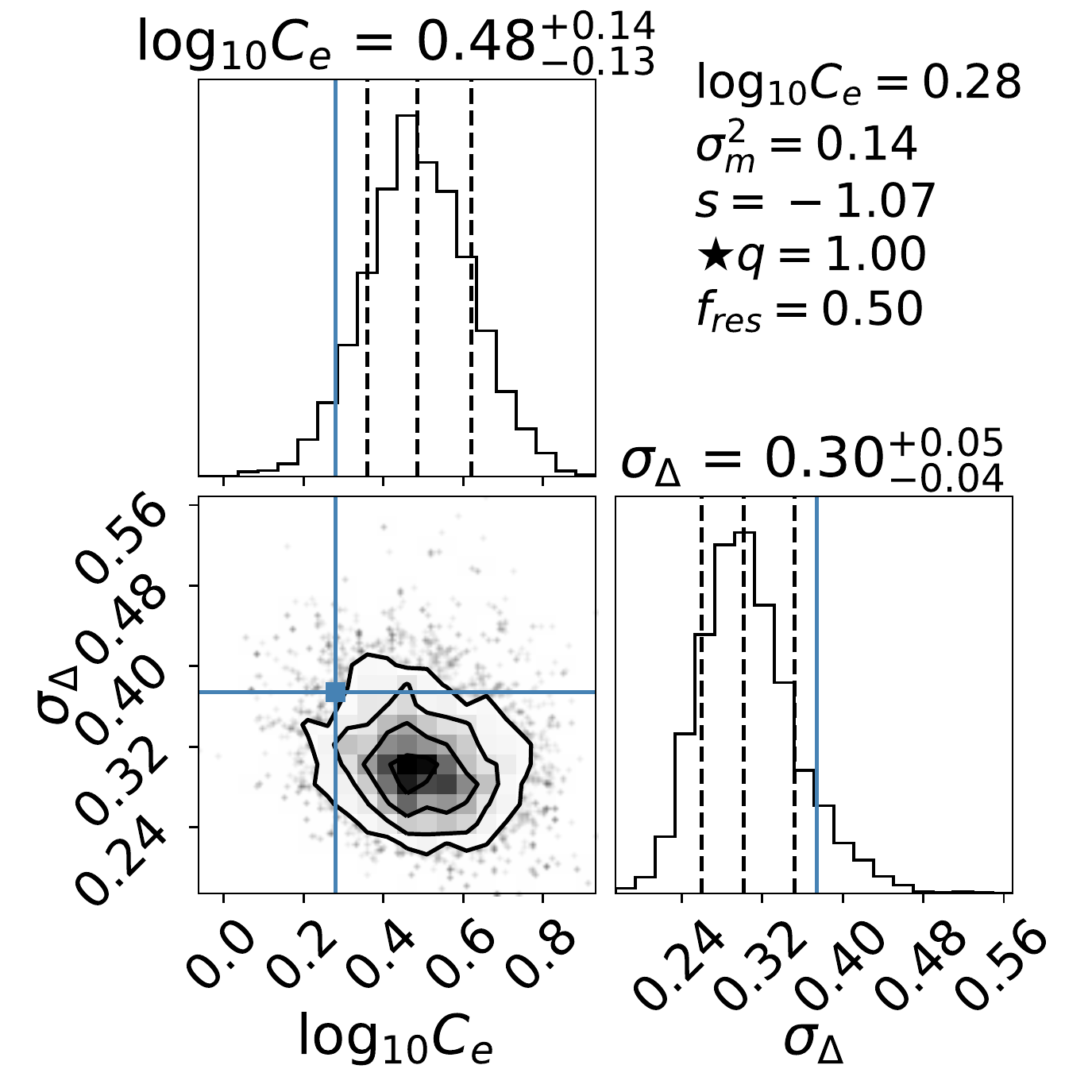}
    \includegraphics[width=0.5\columnwidth]{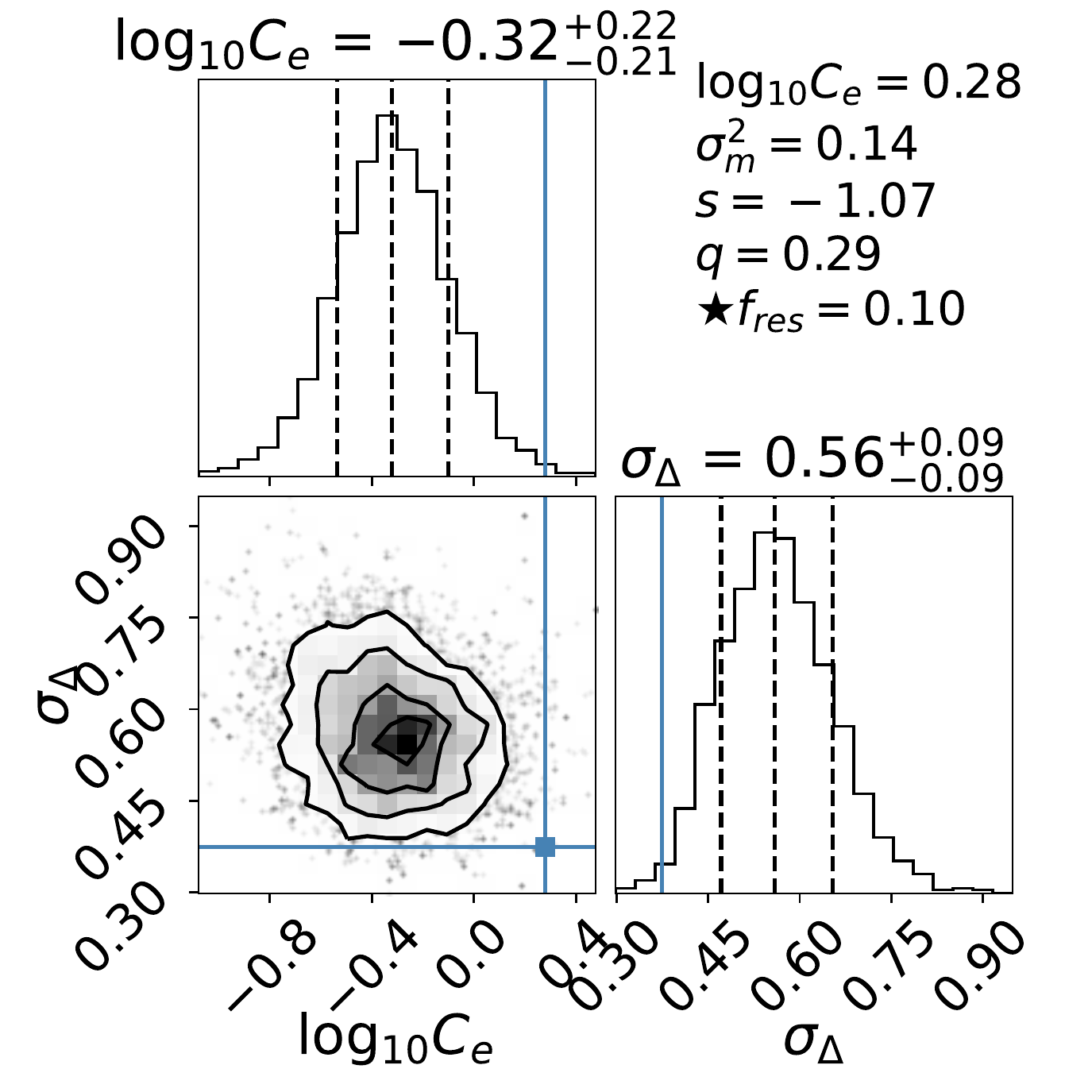}    
    \includegraphics[width=0.5\columnwidth]{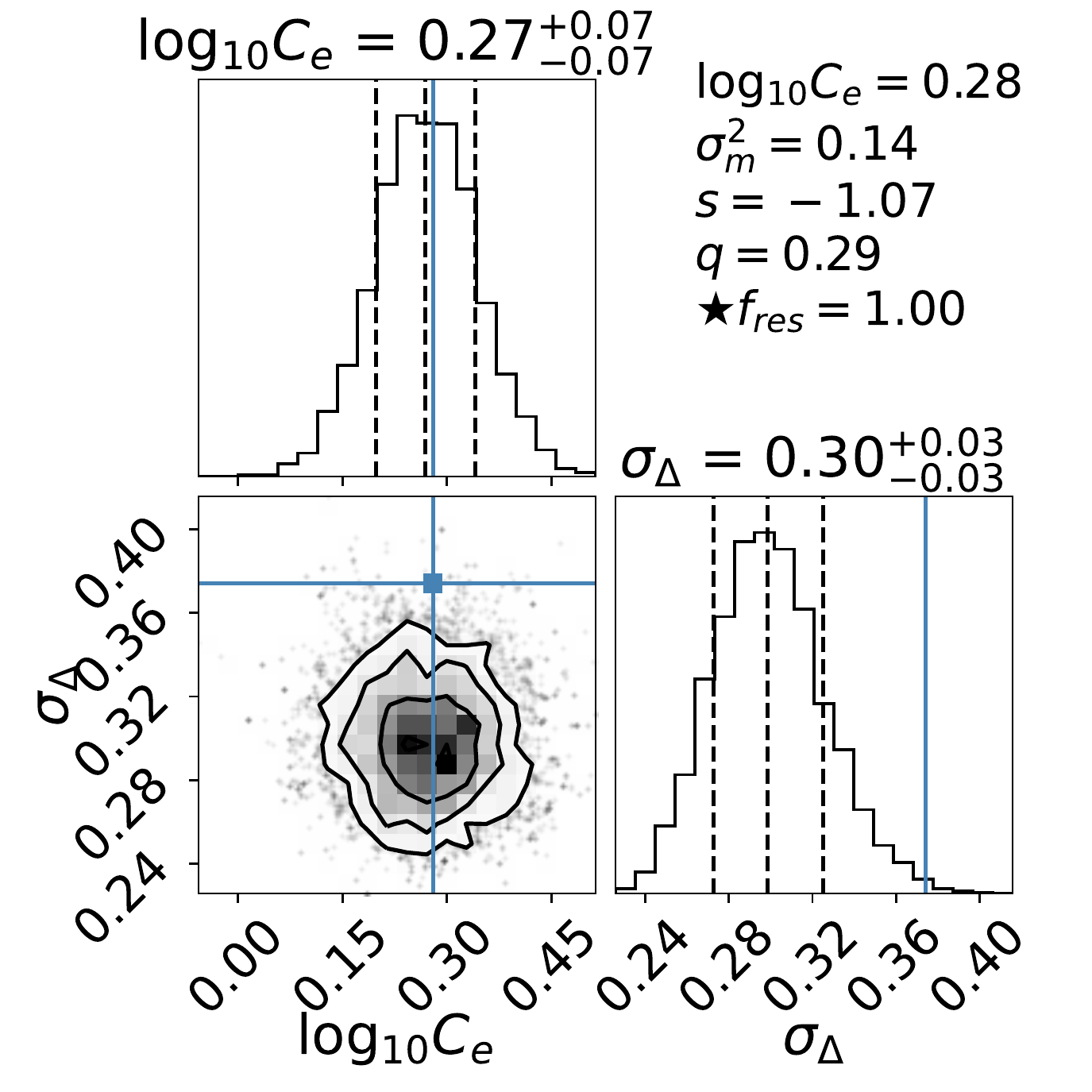}
    
    \caption{Examination of MCMC model performance, fixing the disc structure (the values of $s$ and $q$). The top panel shows the fiducial model parameters. In each panel, all parameters used are labelled in the upper right corner. We then change every single parameter separately (labelled after $\bigstar$) in each panel in the lower half of the figure. We change $\log_{10}C_e$ to $-1$, $\sigma_{m}^2$ to $0.07$ and $0.3$, $s$ to $-0.5$ and $-2$, $q$ to $0$ and $1$ and $f_\mathrm{res}$ to $10\%$ and $100\%$. In each corner plot, blue lines show the true value of $\log_{10}C_e$ and $\sigma_\mathrm{m}$. {The 1$\sigma$ uncertainty is labelled on the top of each column and indicated by left and right dashed lines. The middle dashed lines indicate their median values.} }
    \label{fig:model2_separate}
\end{figure*}

As described in \se{sample}, we make use of planet masses, period ratios, semi-major axes and host masses from the NASA exoplanet database. Given these data as well as disc structure, we can then calculate the exact period ratios if planets are in resonance, and compare them to the observations. In this section, to examine the performance of the MCMC, we generate mock samples by replacing the actual period ratios with those assuming they are in or out of resonance. We randomly select a fraction of planet pairs to be in resonance and the fraction is $f_\mathrm{res}$. If they are not in resonance, the resulting period ratio follows a uniform distribution. If they are in resonance, the resulting period ratio is calculated via \eq{commensurability}. We finally add log-normal noise to the planet masses with $\sigma_m$, which enables us to compare it to the resulting $\sigma_\Delta$. In this way, the mock sample is generated, with known disc parameters. We then run the MCMC model to check whether we can reproduce the input parameters. The mock sample has the same size as the real sample we used in \se{mcmc_analysis}. 

\subsection{Tests without assuming a disc structure}
\label{sec:model2_test_general}
We first examine whether the MCMC can reproduce $\log_{10}(C_eh_\mathrm{1au}^2)$, $s$ and $q$, and whether the fitted $\sigma_\Delta$ is comparable to (the input) $\sigma_m$. Several sets of parameters are used for generating samples and model examination. The default parameters for generating the mock sample are $\log_{10}(C_eh_\mathrm{1au}^2)=-2.94$, $\sigma_m=0.374$, $s=-15/14$, $q=2/7$ and $f_\mathrm{res}=0.5$. We then test different values of $s$, $q$, and $f_\mathrm{res}$. The resulting corner plots are shown in \fg{model2_separate_s}, \fg{model2_separate_q} and \fg{model2_separate_fres}. 
The true value of parameters for generating mock samples is labelled at the top right of each panel and indicated by blue lines. 

Unfortunately, the true values of $\log_{10}(C_eh_\mathrm{1au}^2)$, $s$ and $q$ are not properly retrieved. The posterior distribution gives the expected values for $\log_{10}(C_eh_\mathrm{1au}^2)$, $s$ and $q$, but they are not consistent with the true values. However, the correlation between the parameters is revealed.
The corner plot shows that $s$ negatively correlates with $\log_{10}(C_eh_\mathrm{1au}^2)$, $q$ is positively correlated to $\log_{10}(C_eh_\mathrm{1au}^2)$ and $q$ is positively correlated to $s$. Their correlation is consistent with \eq{equi1} and \eq{equi2}, indicated by green dashed lines. 

Similar to \fg{mcmc_corner_general}, we also plot the posterior distribution of $\log_{10}(\tau_a/\tau_e)$ (\eq{logtate}), which is a rather independent variable. The examination result suggests that $\log_{10}(\tau_a/\tau_e)$ can be fitted within 1.5$\sigma$ error bar in all cases.
The fitted $\sigma_\Delta$ is always very close to the log-normal error $\sigma_m$ we impose for planet mass. It proves that the distribution of resonance offset $\Delta$ resulting from log-normal distributed planet masses also follows a log-normal distribution, when they are in resonance. 

We therefore conclude that our MCMC model is useful for fitting $\log_{10}(\tau_a/\tau_e)$ but not $\log_{10}(C_eh_\mathrm{1au}^2)$, $s$ and $q$ if a disc structure has not been specified. 


\subsection{Tests assuming a disc structure}
\label{sec:model2_test}
We here examine whether the MCMC can reproduce $\log_{10}C_e$, given a disc structure, and whether the fitted $\sigma_\Delta$ is still comparable to $\sigma_m$. We test several sets of parameters. The default parameter set is $\log_{10}C_e=0.28$, $\sigma_m=0.374$, $s=-15/14$, $q=2/7$ and $f_\mathrm{res}=0.5$.
We then change each parameter to two other values, while keeping the other parameters the same. 

The MCMC results are shown in \fg{model2_separate}. Different panels show the fit result for the mock sample generated from different parameter sets, and the true values of the parameters are labelled on the top right and indicated by blue lines. The true value of $\log_{10}C_e$ can always be  reproduced within 1.5$\sigma$ when $f_\mathrm{res}>0.5$. The fitted $\sigma_\Delta$ is always consistent with $\sigma_m$ as well. However, if we decrease the fraction of planets in resonance $f_\mathrm{res}$ to 0.1, the $\log_{10}C_e$ can no longer be retrieved and therefore the MCMC is no longer valid.  However, when fitting the observed data, we have $f_\mathrm{res}\approx0.5$ in most cases (\fg{res_frac} right panel). Therefore the MCMC results for the observed data are reliable.


\bsp	
\label{lastpage}
\end{document}